\documentclass[12pt, a4paper]{article}
 \usepackage[font=small,format=plain,labelfont=bf,up,textfont=normal,up,justification=justified,singlelinecheck=false]{caption}
\usepackage{subcaption}

\usepackage[a4paper, left=2cm,right=2cm]{geometry}
\usepackage[colorlinks=true,linkcolor=black,citecolor=teal,urlcolor=MidnightBlue,filecolor=black]{hyperref}
\usepackage{amsfonts}
\usepackage{amsmath,amssymb}
\usepackage{setspace,braket}
\usepackage[dvipsnames]{xcolor}
\definecolor{SchoolColor}{rgb}{0.6471, 0.1098, 0.1882} 

\usepackage[utf8,applemac]{inputenc}
\usepackage{tensor}
\usepackage{cite}
\usepackage{tikz}
\usepackage{graphicx}
\usepackage{bm} 

\usepackage[utf8,applemac]{inputenc}
\usepackage{tensor}
\usepackage{cite}
\usepackage{tikz}
\usepackage{graphicx}
\graphicspath{{figure/}}
\usepackage{array}
\usepackage{booktabs} 
\usepackage{multirow}

\usepackage{dcolumn}
\usepackage{bm}

\usepackage{verbatim}

\usepackage{textcomp} 
\usepackage{graphicx} 

\numberwithin{equation}{section}
\newcommand{\bea}{\begin{eqnarray}}
\newcommand{\eea}{\end{eqnarray}}
\newcommand{\be}{\begin{equation}}
\newcommand{\ee}{\end{equation}}
\def\nn{\nonumber}

\newcommand{\beqs}{\begin{eqnarray}}
\newcommand{\eeqs}{\end{eqnarray}}
\numberwithin{equation}{section}

\newcommand{\Rmnum}[1]{\uppercase\expandafter{\romannumeral #1\relax}}

\newcommand{\liu}{\color{red}}

\begin{document}
\begin{titlepage}

\begin{flushright}\vspace{-3cm}
{\small
\today }\end{flushright}
\vspace{0.5cm}
\begin{center}
	{{ \LARGE{\bf{Symmetry group at future null infinity \Rmnum{2}:
	
	 Vector  theory}}}}\vspace{5mm}

	\centerline{\large{\bf Wen-Bin  Liu\footnote{liuwenbin0036@hust.edu.cn}, Jiang Long\footnote{
				longjiang@hust.edu.cn}}}
	\vspace{2mm}
	\normalsize
	\bigskip\medskip

	\textit{School of Physics, Huazhong University of Science and Technology, \\ Luoyu Road 1037, Wuhan, Hubei 430074, China
	}
	\vspace{25mm}
	
	\begin{abstract}
		\noindent
		In this paper, we reduce the electromagnetic theory to future null infinity and obtain a vector theory at the boundary. We compute the Poincar\'e flux operators which could be generalized. We quantize the vector theory, and impose normal order on the extended flux operators. It is shown that these flux operators generate the supertranslation and superrotation. When  working out the commutators of these operators, we find that a generalized electromagnetic duality operator should be included as the generators to form a closed symmetry algebra. \end{abstract}
	

\end{center}

\end{titlepage}
\tableofcontents

\section{Introduction}\label{intro}
The study of the gravitational waves led to the famous Bondi-Metzner-Sachs (BMS) group \cite{Bondi:1962px, Sachs:1962wk,Sachs:1962zza} at future null infinity ($\mathcal{I}^+$) in asymptotically flat spacetime. Classically, the BMS group is the semi-direct product of the Lorentz group and supertranslations 
\be 
\text{BMS group}=SO(1,3)\ltimes C^\infty(S^2), 
\ee 
where $C^\infty(S^2)$ denotes the smooth functions on the unit sphere $S^2$.
Supertranslation is an extension of the  global spacetime translation in Minkowski spacetime. It generates an angle-dependent transformation of the retarded time at $\mathcal{I}^+$. 
Over the past decade, the BMS group has been extended to include the so-called superrotation transformations \cite{Barnich:2010eb, Barnich:2009se, Barnich:2010ojg, Barnich:2011mi}.
Just like supertranslation, superrotation is a direct extension of the Lorentz rotation in Minkowski spacetime. The Barnich-Troessaert (BT) superrotations are  generated by local conformal Killing vectors on the celestial sphere  while the Campiglia-Laddha (CL) superrotations \cite{ Campiglia:2014yka, Campiglia:2015yka} are generated by smooth vectors on the celestial sphere. 

 The BMS group plays a central role from the point of view of holography \cite{Strominger:2013jfa,Strominger:2017zoo,Kapec:2016jld,Pasterski:2016qvg,Pasterski:2017kqt,Raclariu:2021zjz,Pasterski:2021rjz,Donnay:2022aba,Donnay:2022wvx} since it is the symmetry group that the boundary field theory should obey. 
 However, the conventional asymptotic symmetry analysis
is  highly sensitive to the fall-off conditions which are imposed on the gravitational waves. Moreover, little is known about the boundary theory in this approach besides the symmetry group. Recently,  
 the BMS group has been identified as the so-called conformal Carroll group of level 2 \cite{Duval_2014a,Duval_2014b,Duval:2014uoa} in the context of Carrollian manifold \cite{Une,Gupta1966OnAA,Henneaux:1979vn}. We may also  study the representation of the BMS group \cite{Chen:2021xkw,Chen:2023pqf} and construct field theories with  Carrollian symmetry \cite{Bagchi:2010zz,Bagchi:2016bcd,Bagchi:2019xfx,Bagchi:2019clu,Banerjee:2020qjj,Hao:2021urq,Henneaux:2021yzg,Bagchi:2022owq,Bagchi:2022xug,Bekaert:2022ipg,Rivera-Betancour:2022lkc,Schwarz:2022dqf,Dutta:2022vkg,Baiguera:2022lsw,Bekaert:2022oeh,Bagchi:2022eav,Saha:2022gjw,Saha:2023hsl,Salzer:2023jqv}.  However, the relation between the Carrollian field theory and the bulk theory is not straightforward in this method. 

 In \cite{Liu:2022mne}, we present a systematic method to overcome these obstacles.  This is achieved by projecting the massless quantum field theory of flat spacetime to $\mathcal{I}^+$ and  constructing the phase space of radiation solution. The outgoing Poincar\'e fluxes are completely determined by the radiation degrees of freedom. We can  generalize the flux operators to define the corresponding supertranslation and superrotation generators. The supertranslation and superrotation form an infinite-dimensional group which could be identified as Carrollian diffeomorphism defined in \cite{Liu:2022mne,Ciambelli:2018xat,Ciambelli:2019lap} classically. The BMS algebra is recovered in the soft limit. Moreover, there is a higher dimensional Virasoro algebra with non-trivial central charge which follows from the time-dependent supertranslations.

In this work, we obtain a vector field theory by projecting the electromagnetic theory  to $\mathcal{I}^+$. The vector field has only two independent propagating degrees of freedom $A_\theta,A_\phi$, where $(\theta,\phi)$ are the spherical coordinates. 
We find the corresponding Poincar\'e flux operators and define the supertranslation and superrotation generators. Interestingly, to make the definition of the superrotation generators reasonable, one should generalize the Lie derivative variation to a covariant variation which is compatible with the metric at $\mathcal{I}^+$. We find the symmetry algebra at $\mathcal{I}^+$ using the generators. In  contrast to the real scalar field theory,  one should include a generalized electromagnetic duality (EM duality) operator $\mathcal{O}_g$ to form a closed symmetry algebra. 

This paper is organized as follows. In section \ref{reviewform} we review the BMS group and explain the terminology used in this paper. In section \ref{fluxessec}, we introduce the ten Poincar\'e  fluxes radiated to $\mathcal{I}^+$ for electromagnetic theory. We quantize the theory at $\mathcal{I}^+$ in the following section. In section \ref{sasec}, we introduce the concept of covariant variation and find a closed Lie algebra by including the new operator $\mathcal{O}_g$. In section \ref{emdualsec}, we interpret the new operator as a generalized EM duality operator. We discuss the antipodal matching condition in section \ref{antipodalsec} and conclude in section \ref{dissec}. Properties for the tensors on the sphere, canonical quantization of the vector field at $\mathcal{I}^+$, details about the calculation of commutators and Green's functions in electromagnetic theory are relegated to several appendices.

\section{Review of the formalism} \label{reviewform}
 In this work, the Minkowski spacetime can be described in Cartesian coordinates $x^\mu=(t,x^i)$, where $\mu=0,1,2,3$ are spacetime coordinates and $i=1,2,3$ are space coordinates. We also use the  retarded coordinate system $(u,r,\theta,\phi)$ 
 and write the Minkowski spacetime as 
 \bea 
 ds^2=-du^2-2du dr+r^2\gamma_{AB}d\theta^Ad\theta^B,\quad A,B=1,2.
 \eea 
  The future null infinity $\mathcal{I}^+$ is a three-dimensional Carrollian manifold with a degenerate metric 
\bea 
ds_{\mathcal{I}^+}^2=\gamma_{AB}d\theta^Ad\theta^B,  \label{degemet}
\eea though the manifold $\mathcal{I}^+$ should be described by three coordinates $(u,\theta^A)$. The spherical coordinates $\theta^A=(\theta,\phi)$ are used to describe the unit sphere $S^2$ whose metric is 
\bea 
\gamma_{AB}=\left(\begin{array}{cc}1&0\\0&\sin^2\theta\end{array}\right).
\eea  
The covariant derivative $\nabla_A$ is  adapted to the metric $\gamma_{AB}$ , while the covariant derivative $\nabla_\mu$ is adapted to the Minkowski metric. 
 Besides the metric \eqref{degemet}, there is also a null vector 
\be 
\bm\chi=\partial_u
\ee which is to generate the retarded time direction. We will use the notation $\dot{F}\equiv \partial_uF$ for any function $F$.

In an asymptotically flat spacetime, transformations  generated by the vector 
\bea 
\xi_f=f\partial_u+\frac{1}{2}\nabla_A\nabla^A f\partial_r-\frac{\nabla^A f}{r}\partial_A+\cdots\label{stxif}
\eea are called supertranslations. 
Similarly, transformations generated by the vector 
\bea 
\xi_Y=\frac{1}{2}u\nabla_AY^A\partial_u-\frac{1}{2}(u+r)\nabla_AY^A\partial_r+(Y^A-\frac{u}{2r}\nabla^A\nabla_BY^B)\partial_A+\cdots\label{srxiY}
\eea  are called superrotations. Usually, the function $f$ and  $Y^A$ are smooth  on $S^2$
\bea 
f=f(\Omega),\quad Y^A=Y^A(\Omega).
\eea 
The metric of $\mathcal{I}^+$ is preserved by  supertranslations
\bea
\delta_{f}\gamma_{AB}&=&0,
\eea while is deformed by  superrotations
\bea 
\delta_{Y}\gamma_{AB}&=&\Theta_{AB}(Y),\label{variationofmetric}
\eea
where
\begin{align}
   \Theta_{AB}(Y)\equiv\nabla_AY_B+\nabla_BY_A-\gamma_{AB}\nabla_C Y^C.
\end{align}
There are six independent global solutions for the conformal Killing equation
\be  
\nabla_AY_B+\nabla_BY_A-\gamma_{AB}\nabla_C Y^C=0
\ee and they correspond to the  Lorentz transformations in Minkowski spacetime. 

The BMS transformation including \eqref{stxif} and \eqref{srxiY} at the Carrollian manifold $\mathcal{I}^+=\mathbb{R}\times S^2$ is generated by \bea 
\xi_{f,Y}=[f(\Omega)+\frac{1}{2}u\nabla_A Y^A(\Omega)]\partial_u+Y^A(\Omega)\partial_A.
\eea Interestingly, the null structure of $\mathcal{I}^+$ is preserved by a more general vector field \cite{Liu:2022mne}
\bea 
\xi_{f,Y}=f(u,\Omega)\partial_u+Y^A(\Omega)\partial_A\label{cardf}
\eea where $f(u,\Omega)$ could depend on the retarded time. The finite transformation corresponding to the vector field \eqref{cardf} is called Carrollian diffeomorphism.


\section{Flux operators}\label{fluxessec}
In this section, we will  study the radiation fluxes of electromagnetic theory. The electromagnetic vector potential is denoted as a $U(1)$ gauge field $a_\mu$. The electric and magnetic fields are combined into an antisymmetric  tensor 
\bea  
f_{\mu\nu}=\partial_\mu a_\nu-\partial_\nu a_\mu.
\eea  More explicitly, the electric field $e_i$ and magnetic field $b_i$ are  
\bea 
e_i=-f_{0i},\quad b_i=\frac{1}{2}\epsilon_{ijk}f^{jk},
\eea where the symbol $\epsilon_{ijk}$ denotes the Levi-Civita tensor in three dimensions. We use the convention $\epsilon_{123}=1$ in Cartesian coordinates.  The action is 
\bea 
S=\int d^4x \sqrt{-g}[-\frac{1}{4}f_{\mu\nu}f^{\mu\nu}+j_\mu a^\mu], \label{action}
\eea where the last term involves a source $j_\mu$ coupled to the  field $a_\mu$ and the source causes the electromagnetic radiation.  
The Maxwell equations are 
\bea 
\nabla_\mu f^{\mu\nu}=-j^\nu,\quad \nabla_{[\mu}f_{\rho\sigma]}=0.\label{maxeq}
\eea 
Usually, the source is located at a finite region of space.  Therefore, we may set it to zero near $\mathcal{I}^+$. 

\subsection{Equations of motion}
To solve the equations \eqref{maxeq}, 
we may impose the fall-off conditions for the vector potential  near $\mathcal{I}^{+}$ 
\bea
a_\mu(t,\bm x)&=&\frac{A_\mu(u,\Omega)}{r}+\sum_{k=2}^\infty \frac{A_\mu^{(k)}(u,\Omega)}{r^k},\quad \mu=0,1,2,3.\label{falloffdicar}
\eea 
In terms of the retarded coordinates,  
\bea 
a_u(u,r,\Omega)&=&\frac{A_u(u,\Omega)}{r}+\sum_{k=2}^\infty \frac{A_u^{(k)}(u,\Omega)}{r^k},\label{falloffau0}\\
a_r(u,r,\Omega)&=&\frac{A_r(u,\Omega)}{r}+\sum_{k=2}^\infty \frac{A_r^{(k)}(u,\Omega)}{r^k},\label{falloffar0}\\
a_A(u,r,\Omega)&=&A_A(u,\Omega)+\sum_{k=1}^\infty \frac{A_A^{(k)}(u,\Omega)}{r^k}.\label{falloffaA0}
\eea We have used the following abbreviations 
\bea 
A_u=A_u^{(1)}, \quad A_r=A_r^{(1)},\quad A_A=A_A^{(0)}.
\eea 
Switching to the retarded coordinates, we find 
\bea 
&&a_u=a_0,\quad a_r=a_0+n_i a_i,\quad a_A=-r Y_A^i a_i.\label{Minkret}
\eea  The vector $n^i$ is the unit normal vector of $S^2$ by embedding it into the Euclidean space $\mathbb{R}^3$, 
\be 
n^i=(\sin\theta\cos\phi,\sin\theta\sin\phi,\cos\theta).
\ee  The vectors $Y_i^A$ are the three strictly CKVs of $S^2$ whose explicit form can be found in Appendix \ref{ckvs}.  They are orthogonal to the normal vector and project the vector to the transverse direction.  The equations \eqref{Minkret} can be solved reversely
\bea 
&&a_0=a_u,\quad a_i=n_i(a_r-a_u)-\frac{1}{r}Y_i^Aa_A.\label{orderbyorder}
\eea This is equivalent to 
\bea 
A_0^{(k)}(u,\Omega)&=&A_u^{(k)}(u,\Omega),\\
A_i^{(k)}(u,\Omega)&=&n_i[A_r^{(k)}(u,\Omega)-A_u^{(k)}(u,\Omega)]-Y_i^AA_A^{(k-1)}(u,\Omega).
\eea 
In this work, we only need the leading  terms
\bea 
A_u(u,\Omega)&=&A_0(u,\Omega),\\
A_r(u,\Omega)&=&A_0(u,\Omega)+n_i A_i(u,\Omega),\\
A_A(u,\Omega)&=&-Y^i_A A_i(u,\Omega)
\eea  and the subleading terms
\bea 
A_u^{(2)}(u,\Omega)&=&A_0^{(2)}(u,\Omega),\\
A_r^{(2)}(u,\Omega)&=&A_0^{(2)}(u,\Omega)+n_i A_i^{(2)}(u,\Omega),\\
A_A^{(1)}(u,\Omega)&=&-Y_A^i A_i^{(2)}(u,\Omega).
\eea In reverse, they are equivalent to 
\bea 
A_0(u,\Omega)&=&A_u(u,\Omega),\label{AoAu}\\
A_i(u,\Omega)&=&n_i[A_r(u,\Omega)-A_u(u,\Omega)]-Y_i^A A_A(u,\Omega)
\eea and 
\bea 
A_0^{(2)}(u,\Omega)&=&A_u^{(2)}(u,\Omega),\\
A_i^{(2)}(u,\Omega)&=&n_i[A_r^{(2)}(u,\Omega)-A^{(2)}_u(u,\Omega)]-Y_i^A A^{(1)}_A(u,\Omega).\label{Ai2Ar2Au2}
\eea 
The electric and magnetic fields are expanded asymptotically 
\bea
e_i(t,\bm x)&=&f_{i0}(t,\bm x)=\frac{E_i(u,\Omega)}{r}+\sum_{k=2}^\infty \frac{E_i^{(k)}(u,\Omega)}{r^k},\\
b_i(t,\bm x)&=&\frac{1}{2}\epsilon_{ijk}f_{jk}(t,\bm x)=\frac{B_i(u,\Omega)}{r}+\sum_{k=2}^\infty \frac{B_i^{(k)}(u,\Omega)}{r^k}
\eea 
where \bea E_i(u,\Omega)&=&-n_i\dot A_0-\dot A_i,\label{EiA0}\\
E_i^{(2)}(u,\Omega)&=&-Y_i^A\partial_AA_0(u,\Omega)-n_i A_0(u,\Omega)-n_i \dot{A}_0^{(2)}(u,\Omega)-\dot{A}_i^{(2)}(u,\Omega),\\
B_i(u,\Omega)&=&-\epsilon_{ijk}n_j\dot{A}_k,\\
B_i^{(2)}(u,\Omega)&=&-\epsilon_{ijk}Y_j^A\partial_A A_k-\epsilon_{ijk}n_j A_k-\epsilon_{ijk}n_j\dot{A}_k^{(2)}.\label{Bi2Ai}
\eea Using the relation \eqref{AoAu}-\eqref{Ai2Ar2Au2}, they are
\bea 
E_i(u,\Omega)&=&Y_i^A\dot{A}_A-n_i\dot{A}_r,\\
B_i(u,\Omega)&=&-\widetilde{Y}_i^A\dot{A}_A,\\
E_i^{(2)}(u,\Omega)&=&-Y_i^A(\partial_A A_u^{}-\dot{A}_A^{(1)})-n_i(A_u^{}+\dot{A}_r^{(2)}),\\
B_i^{(2)}(u,\Omega)&=&{\widetilde{Y}}_i^A(\partial_A A_u^{}-\dot{A}_A^{(1)}-\partial_A A_r^{})+n_i\epsilon^{AB}\partial_A A_B.
\eea The vector $\widetilde{Y}_i^A$ is related to the Killing vector $Y_{jk}^A$ defined in Appendix \ref{ckvs} by \be \widetilde{Y}_i^A=\frac{1}{2}\epsilon_{ijk}Y_{jk}^A.\ee  In the following, we will also call  $E_i, E_i^{(2)},\cdots$ the electric fields and  $B_i,B_i^{(2)},\cdots$ the magnetic fields at $\mathcal{I}^+$. Since the electric field $e_i$ and magnetic field $b_i$ are gauge invariant under gauge transformations, the electric fields $E_i,E_i^{(2)},\cdots$ and magnetic fields $B_i,B_i^{(2)},\cdots$  { are} also gauge invariant quantities at $\mathcal{I}^+$. 
Using the following properties 
\begin{align}
    n_iY^A_i=0,\qquad n_i\widetilde{Y}^A_i=0,\qquad Y_i^A\widetilde Y^B_i=0,
\end{align}
we find that the leading electric field is orthogonal on shell to the leading magnetic field 
\begin{align}
    E_i(u,\Omega)B_i(u,\Omega)=0.
\end{align}
The same is true for the subleading order fields $E_i^{(2)}(u,\Omega)$ and $B_i^{(2)}(u,\Omega)$.

We can also write the electromagnetic fields in retarded coordinates 
\bea 
f_{ur}&=&\frac{\dot{A}_r}{r}+\frac{A_u^{}+\dot A_r^{(2)}}{r^2}+\cdots,\\
f_{uA}&=&\dot{A}_A+\frac{-\partial_A A_u^{}+\dot{A}_A^{(1)}}{r}+\frac{-\partial_A A_u^{(2)}+\dot{A}_A^{(2)}}{r^2}+\cdots,\\
f_{rA}&=&-\frac{\partial_A A_r}{r}+\frac{-A_A^{(1)}-\partial_A A_r^{(2)}}{r^2}+\cdots,\\
f_{AB}&=&\partial_A A_B-\partial_B A_A+\frac{\partial_A A_B^{(1)}-\partial_B A_A^{(1)}}{r}+\frac{\partial_A A_B^{(2)}-\partial_B A_A^{(2)}}{r^2}+\cdots.
\eea 
From the equation of motion, we find 
\bea \dot A_r^{}&=&0,\label{eqn2}\\
\ddot{A}_r^{(2)}+\dot{A}_u^{}&=&\gamma^{AB}\nabla_A\dot{A}_B\label{eqn1}.
\eea  The first equation \eqref{eqn2} fixes the radial component of the vector potential up to a time-independent function
\be 
A_r=\varphi_1(\Omega).
\ee For the second equation \eqref{eqn1}, we may  integrate it 
\bea 
A_u+\dot{A}_r^{(2)}=\nabla_A{A}^A+\varphi(\Omega)\label{Ar(2)},
\eea where $\varphi(\Omega)$ is also time-independent. 
Therefore, the on-shell electric and magnetic fields are 
\bea 
E_i(u,\Omega)&=&Y_i^A\dot{A}_A,\\
B_i(u,\Omega)&=&-\widetilde{Y}_i^A\dot{A}_A,\\
E_i^{(2)}(u,\Omega)&=&-Y_i^A(\partial_A A_u-\dot{A}_A^{(1)})-n_i[\nabla_A A^A+\varphi(\Omega)],\\
B_i^{(2)}(u,\Omega)&=&\widetilde{Y}_i^A[\partial_A A_u^{}-\dot{A}_A^{(1)}-\partial_A  \varphi_1(\Omega)]+n_i\epsilon^{AB}\partial_A A_B.
\eea 
Using the orthogonality relation in Appendix \ref{ckvs}, the electric {field} $E_i$ and magnetic field $B_i$ are orthogonal to the normal vector of $S^2$ at $\mathcal{I}^+$
\be 
E_i n_i=B_i n_i=0.
\ee  On the other hand, at the subleading order, the electric field $E_i^{(2)}$ and magnetic field $B_i^{(2)}$ are not orthogonal to the normal vector $n_i$. We may define their radial components
\bea 
E^{(2)}(u,\Omega)&=&n_i E_i^{(2)}=-\nabla_A A^A(u,\Omega)-\varphi(\Omega),\\
B^{(2)}(u,\Omega)&=&n_iB_i^{(2)}=\epsilon^{AB}\partial_AA_B
\eea { for later convenience.}

\subsection{Poincar\'e fluxes}
The electromagnetic theory is invariant under Poincar\'e transformations, so there are ten corresponding Poincar\'e fluxes which are related to the conservation laws
\bea 
\partial_\mu T^{\nu\mu}=0,\quad \partial_\rho M^{\mu\nu\rho}=0,
\eea where 
\be 
M^{\mu\nu\rho}=x^\mu  T^{\nu\rho}-x^\nu T^{\mu\rho}.
\ee The stress-energy tensor is 
\bea 
T_{\mu\nu}=f_{\mu\rho}f_{\nu}^{\ \rho}-\frac{1}{4}\eta_{\mu\nu} f_{\rho\sigma}f^{\rho\sigma}.
\eea It is easy to derive the following ten Poincar\'e fluxes. 
\begin{itemize}
\item Energy flux
\bea 
\frac{dP^0}{du}&=&-\int d{S}_iT^{0i}\nn\\&=&-\int_{S^2} d\Omega n_i\epsilon_{ijk}E_jB_k\nn\\&=&-\int_{S^2}d\Omega \dot{A}_A\dot{A}^A.
\eea 
\item Momentum fluxes
\bea 
\frac{dP^i}{du}&=&-\int dS_jT^{ji}\nn\\&=&-\frac{1}{2}\int_{S^2} d\Omega n_i(E_j^2+B_j^2)\nn\\&=&-\int_{S^2}d\Omega n_i \dot{A}_A\dot{A}^A.
\eea 
\item Angular momentum fluxes 
\bea 
\frac{dL^{ij}}{du}&=&-\int dS_k M^{ijk}\nn\\&=&\int_{S^2}d\Omega [E^{(2)}(n_iE_j-n_j E_i)+B^{(2)}(n_iB_j-n_jB_i)]\nn\\ 
&=&\int_{{S^2}} d\Omega \dot{A}_C\nabla_AA_B(\gamma^{AB}Y_{ij}^C+\gamma^{BC}Y_{ij}^A-\gamma^{CA}Y_{ij}^B).
\eea At the last step, we have used the definition of $E^{(2)}$ and discarded the term $\varphi(\Omega)$ since it does not affect the total angular momentum radiated to null infinity.
\item Center-of-mass fluxes
\bea 
\frac{dL^{0i}}{du}&=&-\int dS_j M^{0ij}\nn\\&=&-\frac{1}{2}u\int_{S^2} d\Omega  n_i(E_j^2+B_j^2)\nn\\&&+\int_{S^2}d\Omega [E_iE^{(2)}+B_iB^{(2)}+n_i n_j \epsilon_{jmn}(E_mB_n^{(2)}+E^{(2)}_m B_n)-n_i(E_mE_m^{(2)}+B_m  B_m^{(2)})]\nn\\
&=&-u\int_{S^2} d\Omega n_i \dot A_A\dot A^A+\int_{S^2} d\Omega  \dot{A}_C\nabla_AA_B(-\gamma^{AB}Y_{i}^C-\gamma^{BC}Y_{i}^A+\gamma^{CA}Y_{i}^B).
\eea At the last step, we have discarded the term involving $\varphi(\Omega)$  with the same reason as the case of the angular momentum fluxes. 
\end{itemize}
From the Poincar\'e fluxes, we read out the energy flux density operator 
\be 
T(u,\Omega)=\dot A_A\dot A^A,
\ee and the angular momentum flux density operator 
\be 
M_A(u,\Omega)=\frac{1}{2}(\dot A^B\nabla^C A^D-A^B\nabla^C\dot A^D) P_{ABCD}, \label{angudensity}
\ee where we have defined a rank 4 tensor  
\bea 
P_{ABCD}=\gamma_{AB}\gamma_{CD}+\gamma_{AC}\gamma_{BD}-\gamma_{AD}\gamma_{BC}.
\eea Useful properties of the tensor $P_{ABCD}$ are collected in Appendix \ref{rank4subsec}.

The data of the fluxes depend on angular directions as well as retarded time. 
Therefore, we may construct two smeared operators on $\mathcal{I}^+$
\bea 
\mathcal{T}_f&=&\int du d\Omega f(u,\Omega)T(u,\Omega),\label{energyfluxes}\\
\mathcal{M}_Y&=&\int du d\Omega Y^A(u,\Omega)M_A(u,\Omega)\label{angularmomentumfluxes}
\eea without losing any information. Classically, the two operators are related to the fluxes radiated to $\mathcal{I}^+$. More explicitly, 
\begin{itemize}
\item When $f={ -}\theta(u_0-u)$, \eqref{energyfluxes} is the energy radiated to $\mathcal{I}^+$ from $u=-\infty$ to $u=u_0$.
\item When $f={ -}\theta(u_0-u)n^i$, \eqref{energyfluxes} is the {$i$}-th component of the momentum radiated to $\mathcal{I}^+$ from $u=-\infty$ to $u=u_0$. 
\end{itemize}  Obviously, $\mathcal{T}_f$ should be regarded as a generalized Fourier transformation of the energy flux density $T(u,\Omega)$ on $\mathcal{I}^+$. It encodes the same information as the energy flux density operator when $f$ can be any smooth function on $\mathcal{I}^+$.
The test vector function $Y^A$ in  $\mathcal{M}_Y$ can also depend on the retarded time. Note that the test functions $f$ and $Y^A$ here are not related to supertranslations and superrotations defined in the context of asymptotic symmetry analysis so far. However,
we may use the terminology in \cite{Liu:2022mne} and distinguish the following four cases 
\bea  
\text{Special supertranslation (SST)}&\Leftrightarrow&\dot{f}=0,\\
\text{General supertranslation (GST)}&\Leftrightarrow&\dot{f}\not=0,\\
\text{Special superrotation (SSR)}&\Leftrightarrow&\dot{Y}^A=0,\\
\text{General superrotation (GSR)}&\Leftrightarrow&\dot{Y}^A\not=0.
\eea 

Note that there is an ambiguity in the definition of the angular momentum flux density operator \eqref{angudensity}. To illustrate this problem, we may define  a family of angular momentum flux density operators
\bea 
{M}_A(\lambda)=[\lambda\dot A^B\nabla^CA^D-(1-\lambda)A^B\nabla^C\dot A^D]P_{ABCD}\label{MYlocal}
\eea and the corresponding smeared operator 
\be 
\mathcal{M}_Y(\lambda)\equiv \int du d\Omega Y^A(u,\Omega)M_A(\lambda).\label{MYsm}
\ee To reproduce the angular momentum fluxes, we set $Y^A $ to be a Killing vector of $S^2$ which is independent of $u$. Using integration by parts the smeared operator $\mathcal{M}_Y(\lambda)$ is independent of $\lambda$. The one-parameter family of the operators \eqref{MYsm} shares the same classical meaning. We will fix the choice of $\lambda$ in the next section.



\section{Quantization}\label{canosec}
In the previous section, we found the Poincar\'e fluxes at $\mathcal{I}^+$.  The densities 
$T(u,\Omega), M_A(u,\Omega)$ are classical objects so far. In this section, we will  use the covariant phase space method\cite{Lee:1990nz,Wald:1999wa} to quantize  the densities $T(u,\Omega)$ and $M_A(u,\Omega)$. For simplicity, we will use the radial gauge $a_r=0$.  

 The variation of the action \eqref{action} is given by a bulk term which is proportional to the equation of motion, and a boundary term 
\bea 
\delta S=-\int_{\text{bdy}} (d^3x)_\mu f^{\mu}_{\ \nu}\delta a^\nu+\int_{\text{bulk}}d^4x (\text{EOM})
\eea where the volume form $(d^3x)_\mu$ is 
\be 
(d^3x)_\mu=\frac{1}{6}\epsilon_{\mu\nu\rho\sigma}dx^\nu\wedge dx^\rho\wedge dx^\sigma.
\ee The presymplectic potential form is 
\bea 
\bm\Theta(\delta a;a)=-(d^3x)^\mu f_{\mu\nu}\delta a^\nu.
\eea 
We can obtain the presymplectic form 
\bea 
\bm\omega(\delta_1 a,\delta_2 a;a)=\delta_1\bm\Theta(\delta_2a;a)-\delta_2\bm\Theta(\delta_1a;a)=-(d^3x)^\mu [\delta_1 f_{\mu\nu}\delta_2 a^\nu-(1\leftrightarrow 2)].
\eea 
Using the fall off condition \eqref{falloffau0}-\eqref{falloffaA0}, 
the  presymplectic form becomes
\bea 
\bm\omega(\delta_1 a,\delta_2 a;a)= -\sin\theta du\wedge d\theta\wedge d\phi [\delta_1 \dot A_A \delta_2 A^A-(1\leftrightarrow 2)]+\mathcal{O}(r^{-1}).
\eea 
Now it is straightforward to work out the fundamental commutators at $\mathcal{I}^+$
\bea 
\ [A_A(u,\Omega),A_{B}(u',\Omega')]&=&\frac{i}{2}\gamma_{AB}\alpha(u-u')\delta(\Omega-\Omega'),\label{AA}\\
\ [A_{A}(u,\Omega),\dot A_{B}(u',\Omega')]&=&\frac{i}{2}\gamma_{AB}\delta(u-u')\delta(\Omega-\Omega'),\label{AdotA}\\
\ [\dot A_{A}(u,\Omega),\dot A_{B}(u',\Omega')]&=&\frac{i}{2}\gamma_{AB}\delta'(u-u')\delta(\Omega-\Omega'),\label{dotAdotA}
\eea 
where the Dirac function on the sphere reads out explicitly as
\be 
\delta(\Omega-\Omega')=\frac{1}{\sin\theta}\delta(\theta-\theta')\delta(\phi-\phi'),\label{diracsphere}
\ee 
and the function $\alpha(u-u')$ is defined as
\bea 
\alpha(u-u')=\frac{1}{2}[\theta(u'-u)-\theta(u-u')].
\eea 
 The commutators \eqref{AA}-\eqref{dotAdotA} have already been found in the literature \cite{Ashtekar:1981bq,Ashtekar:1981sf,Ashtekar:1987tt,Strominger:2017zoo}. In Appendix \ref{canoquan}, we use the standard canonical quantization method \cite{1995iqft.book.....P} to obtain the same answer. Similar to the scalar case, we find the following correlators
\bea 
\langle 0|A_A(u,\Omega)A_B(u',\Omega')|0\rangle&=&\gamma_{AB}\beta(u-u')\delta(\Omega-\Omega'),\\
\langle 0|A_A(u,\Omega)\dot A_B(u',\Omega')|0\rangle&=&\gamma_{AB}\frac{1}{4\pi(u-u'-i\epsilon)}\delta(\Omega-\Omega'),\\
\langle 0|\dot A_A(u,\Omega) A_B(u',\Omega')|0\rangle&=&-\gamma_{AB}\frac{1}{4\pi(u-u'-i\epsilon)}\delta(\Omega-\Omega'),\\
\langle 0|\dot A_A(u,\Omega)\dot A_B(u',\Omega')|0\rangle&=&-\gamma_{AB}\frac{1}{4\pi(u-u'-i\epsilon)^2}\delta(\Omega-\Omega').
\eea  We have defined a divergent function
\bea 
\beta(u-u')=\int_0^\infty \frac{d\omega}{4\pi\omega}e^{-i\omega(u-u'-i\epsilon)}.
\eea 
From this divergent function, we could find a finite result by considering the following difference
\bea 
\beta(u-u')-\beta(u'-u)=\frac{i}{2}\alpha(u-u').\label{alphabeta}
\eea  For more details of the function $\beta(u-u')$, we refer readers to \cite{Liu:2022mne}. 

After quantization, the densities $T(u,\Omega),M_{A}(u,\Omega)$ are quantum operators. We may refine their definition by using normal ordering 
\bea 
T(u,\Omega)&=&:\dot A_A(u,\Omega)\dot A^A(u,\Omega):,\\
M_A(u,\Omega)&=&\frac{1}{2}:(\dot A^B\nabla^C A^D-A^B\nabla^C\dot A^D): P_{ABCD}.
\eea Now the vacuum expectation values of these flux operators become zero
\bea 
\langle 0|T(u,\Omega)|0\rangle=\langle0|M_A(u,\Omega)|0\rangle=0.
\eea Using the normal ordering, we find the following two-point functions
\bea 
\langle 0|T(u,\Omega)T(u',\Omega')|0\rangle&=&\frac{\delta^{(2)}(0)}{4\pi^2(u-u'-i\epsilon)^4}\delta(\Omega-\Omega'),\label{TT2pt}\\
\langle 0|T(u,\Omega)M_{A'}(u',\Omega')|0\rangle&=&0,\label{TR2pt}\\
\langle 0|M_{A}(u,\Omega)M_{B'}(u',\Omega')|0\rangle&=&-\frac{\beta(u-u')-\frac{1}{4\pi}}{4\pi(u-u'-i\epsilon)^2}\Lambda^{(1)}_{AB'}(\Omega,\Omega').\label{MM2pt}
\eea The two-point functions have similar structure as those in the scalar theory. The divergent constant $\delta^{(2)}(0)$ is the Dirac function \eqref{diracsphere}  on the sphere with the argument equalling to 0. The tensor  $\Lambda^{(1)}_{AB'}$ is 
 \bea 
  \Lambda^{(1)}_{AB'}(\Omega,\Omega')&=&\frac{1}{2}P_{ABCD}P_{B'F'G'H'}[\gamma^{BF'}\gamma^{DH'}\delta(\Omega-\Omega')\nabla^C\nabla^{G'}\delta(\Omega-\Omega')\nn\\&&
 -\gamma^{BH'}\gamma^{DF'}\nabla^C\delta(\Omega-\Omega')\nabla^{G'}\delta(\Omega-\Omega')].
\eea  We use the subscript ${}^{(1)}$  to distinguish from the tensor $\Lambda$ which has been defined in the scalar theory. 
The vanishing of the two-point function \eqref{TR2pt} indicates that $M_A$ is orthogonal to $T$. This corresponds to the operator \eqref{MYlocal} with \be  \lambda=\frac{1}{2}.
\ee For any other value of  $\lambda$, the energy flux density operator is not orthogonal to the angular momentum flux density operator. The orthogonality condition fixes the value of $\lambda$ uniquely.

\section{Symmetry algebra at \texorpdfstring{$\mathcal{I}^+$}{}}\label{sasec}
In this section, we first obtain the variation of the field $A_A$ generated by supertranslation and superrotation. The variation from superrotation is not compatible with the metric of $\mathcal{I}^+$. This motivates us to  define a covariant variation of $A_A$ under superrotation so that we can  identify the  operators $\mathcal{T}_f$ and $\mathcal{M}_Y$ as supertranslation and superrotation generators, respectively. The symmetry algebra can be found in the last part of this section. 

\subsection{Covariant variation}
 The Lie derivative of the (co-)vector field $a_\mu$ is
\bea 
\mathcal{L}_\xi a_\mu=\xi^\nu\partial_\nu a_\mu+a_\nu\partial_\mu \xi^\nu. 
\eea From the fall-off conditions 
\bea  
a_u(t,\bm x)&=&\frac{A_u(u,\Omega)}{r}+\mathcal{O}\left(\frac{1}{r^2}\right),\label{fallau}\\
a_A(t,\bm x)&=&A_A(u,\Omega)+\frac{A_A^{(1)}(u,\Omega)}{r}+\mathcal{O}\left(\frac{1}{r^2}\right),\label{fallaA}
\eea 
we can find the variation of the radiation degrees of freedom $A_A$ on $\mathcal{I}^+$. The result is collected as follows.
\begin{itemize}
\item When $\xi=\xi_f$, the supertranslation variation of the vector field  $A_A$ is
\bea
\delta_fA_A=f\dot A_A.\label{stdiffeo}
\eea  
\item When $\xi=\xi_Y$, the superrotation variation of the vector field $A_A$ is 
\bea 
\delta_Y A_A&=&\frac{1}{2}u\nabla_CY^C \dot A_A+Y^C\nabla_CA_A+A_C\nabla_AY^C.
\eea 
\end{itemize}
Now the Lie derivative of the vector field $a^\mu$ is 
\be 
\mathcal{L}_\xi a^\mu=\xi^\nu\partial_\nu a^\mu-a^\nu\partial_\nu\xi^\mu.
\ee From the fall-off conditions 
\bea  
a^u(t,\bm x)&=&{ 0},\label{falloffau}\\
a^r(t,\bm x)&=&-\frac{A_u(u,\Omega)}{r}+\mathcal{O}\left(\frac{1}{r^2}\right),\label{falloffar}\\
a^A(t,\bm x)&=&\frac{A^A(u,\Omega)}{r^2}+\mathcal{O}\left(\frac{1}{r^3}\right)
\label{falloffaA}\eea  
where 
\bea 
A^A(u,\Omega)=\gamma^{AB}A_A(u,\Omega),\eea 
we can also find the variation of  $A^A$ on $\mathcal{I}^+$
\bea  
\delta_fA^A&=&f\dot A^A,\\
\delta_YA^A&=&\frac{1}{2}u\nabla_BY^B\dot{A}^A+Y^C\nabla_CA^A-A^C\nabla_CY^A+A^A\nabla_BY^B.
\eea
 The supertranslation is compatible with the metric
\bea 
\delta_f A^A&=&\gamma^{AB}\delta_f A_B.\label{covfA}
\eea However, the superrotation variation of the field $A_A$ is not covariant since its indices cannot be raised or lowered by the metric of $S^2$
\bea  \delta_Y A^A&\not=&\gamma^{AB}\delta_Y A_B.\label{covYA}
\eea This is expectable since the metric of $S^2$ is not invariant even for SSRs, which has been shown in equation \eqref{variationofmetric}.

Now we try to define a covariant variation $\delta\hspace{-6pt}\slash$ for the  field on $\mathcal{I}^+$. This is denoted by 
\bea 
\delta\hspace{-6pt}\slash_f (\cdots)\eea for supertranslations and \bea  \delta\hspace{-6pt}\slash_Y (\cdots)
\eea for superrotations. The $\cdots$  in parenthesis is any well-defined field on $\mathcal{I}^+$. For example, the supertranslation and superrotation  for the scalar field $\Sigma$ in \cite{Liu:2022mne} are 
\be 
\delta\hspace{-6pt}\slash_f \Sigma=f\dot\Sigma\quad\text{and}\quad \delta\hspace{-6pt}\slash_Y \Sigma=\frac{1}{2}u\nabla_BY^B\dot\Sigma+Y^A\nabla_A\Sigma+\frac{1}{2}(\nabla_BY^B) \Sigma,
\ee respectively. For the vector field, these are denoted as 
\bea 
\delta\hspace{-6pt}\slash_f A_A, \quad \delta\hspace{-6pt}\slash_Y A_A,\quad \delta\hspace{-6pt}\slash_f A^A\quad\text{and}\quad \delta\hspace{-6pt}\slash_Y A^A.
\eea 
We use a slash to distinguish it from the original variation induced by Lie derivative. From \eqref{covfA}, there is no need to modify the variation of the vector field under supertranslation
\be 
\delta\hspace{-6pt}\slash_f(\cdots)=\delta_f(\cdots).
\ee For superrotations, the covariant variation should satisfy the following conditions 
\begin{itemize}
\item Linearity. For any vector fields $Y^A$ and $Z^A$ and any constants $c_1,c_2$, 
\bea 
\delta\hspace{-6pt}\slash_{c_1 Y+c_2 Z}(\cdots)=c_1\delta\hspace{-6pt}\slash_{Y}(\cdots)+c_2\delta\hspace{-6pt}\slash_{Z}(\cdots).
\eea Also, for any two fields $F_1$ and $F_2$ of the same type, we require 
\be 
\delta\hspace{-6pt}\slash_Y (F_1+F_2)=\delta\hspace{-6pt}\slash_Y F_1+\delta\hspace{-6pt}\slash_Y F_2.
\ee 
\item Leibniz rule. For any two fields $F_1$ and $F_2$ on $\mathcal{I}^+$, their tensor product should obey the Leibniz rule
\bea  
\delta\hspace{-6pt}\slash_Y(F_1F_2)=F_2\delta\hspace{-6pt}\slash_YF_1+F_1\delta\hspace{-6pt}\slash_YF_2.
\eea 
\item Metric compatibility. The covariant variation of the metric should be zero
\be 
\delta\hspace{-6pt}\slash_Y \gamma_{AB}=0.\label{metriccompa}
\ee 
\item For the scalar field $\Sigma$, the variation is the variation induced by  Lie derivative
\be 
\delta\hspace{-6pt}\slash_Y\Sigma=\delta_Y\Sigma.\label{vascalar}
\ee 
\end{itemize}
From the linearity condition, a possible definition of $\delta\hspace{-6pt}\slash_YA_A$ may be 
\be 
\delta\hspace{-6pt}\slash_Y A_A=\delta_Y A_A-\Gamma_A^{\ C}(Y)A_C.\label{conn}
\ee The rank 2 tensor $\Gamma_A^{\ C}(Y)$ should be linear in $Y$ and independent of $A_A$.  Now using the Leibniz rule and the condition \eqref{vascalar}, we should find 
\be 
\delta\hspace{-6pt}\slash_Y A^A=\delta_Y A^A+\Gamma^{\ A}_{C}(Y)A^C\ee and 
\bea 
\delta\hspace{-6pt}\slash_Y\gamma_{AB}=\delta_Y\gamma_{AB}-\Gamma_A^{\ C}(Y)\gamma_{CB}-\Gamma_B^{\ C}(Y)\gamma_{AC}=\Theta_{AB}(Y)-\Gamma_{AB}(Y)-\Gamma_{BA}(Y).
\eea 
We have defined $\Gamma_{AB}(Y)$ with lower indices as 
\bea 
\Gamma_{AB}(Y)=\gamma_{CB}\Gamma_A^{\ C}(Y).
\eea 
The metric compatibility condition \eqref{metriccompa} implies \be   \Gamma_{AB}(Y)+\Gamma_{BA}(Y)=\Theta_{AB}(Y).\label{compa}
\ee We decompose the connection into symmetric and antisymmetric part 
\bea 
\Gamma_{AB}(Y)=\Gamma_{(AB)}(Y)+\Gamma_{[AB]}(Y).
\eea The metric compatibility condition fixes the symmetric part 
\be 
\Gamma_{(AB)}(Y)=\frac{1}{2}\Theta_{AB}(Y).
\ee The antisymmetric part  should be proportional to the Levi-Civita tensor $\epsilon_{AB}$ 
\be 
\Gamma_{[AB]}(Y)=\epsilon_{AB}\Upsilon(Y)
\ee where $\Upsilon(Y)$ is an arbitrary linear function of $Y$. To remove this ambiguity, we may require the connection $\Gamma_{AB}(Y)$ to be symmetric 
\be 
\Gamma_{AB}(Y)=\Gamma_{BA}(Y).
\ee Then the connection is uniquely fixed to 
\be 
\Gamma_{AB}(Y)=\frac{1}{2}\Theta_{AB}(Y).\label{connection}
\ee 
The symmetric connection is a rank 2 traceless tensor. Therefore, one can use the metric $\gamma_{AB}$ to raise and lower its indices. For example, 
\bea 
\Gamma^{AB}(Y)=\gamma^{AC}\Gamma_{C}^{\ B}(Y).
\eea 
In the following, we will use the covariant variation whose connection is symmetric.  As a consequence of the definition, we find 
\bea 
\delta\hspace{-6pt}\slash_Y \gamma^{AB}=0,\quad \delta\hspace{-6pt}\slash_Y\epsilon_{AB}=0.
\eea 
Due to the nice property of the covariant variation, we may regard the transformation $\delta\hspace{-6pt}\slash_YA_A$ as the ``real'' superrotation of the vector field. The variation $\delta_YA_A$ induced by diffeomorphisms is only partial variation of the superrotations. 

\subsection{Supertranslation and superrotation generators}\label{myaa}
There is another variation  defined by the commutators
\bea  
\ [\mathcal{T}_f,A_{A'}(u',\Omega')]&=&-if(u',\Omega')\dot A_{A'}(u',\Omega'),\label{supertranslationge}\\
\ [\mathcal{M}_Y,A_{A'}(u',\Omega')]&=&-i\Delta_{A'}(Y;A;u',\Omega')+\frac{i}{2}\int du \alpha(u-u')\Delta_{A'}(\dot Y;A;u,\Omega')\label{superrotationge}
\eea  
where \bea 
\Delta_A(Y;A;u,\Omega)=Y^B(u,\Omega)\nabla^DA^C(u,\Omega)\rho_{BCDA}(\Omega)+\frac{1}{2}A^D(u,\Omega)\nabla^BY^C(u,\Omega)P_{BCAD}(\Omega).\nn\\
\eea  The rank 4 tensor $\rho_{BCDA}$ is defined as
\bea 
\rho_{BCDA}=\frac{1}{2}(P_{ABCD}+P_{ADCB})=\gamma_{BD}\gamma_{CA}.
\eea 
Interestingly, \eqref{supertranslationge} is exactly the variation of the vector field $A_A$ under the supertranslation 
\eqref{stdiffeo} up to a constant factor. Moreover, after some algebra, we could find 
\bea
\Delta_A(Y;A;u,\Omega)&=&Y^C\nabla_C A_A+A_C\nabla_AY^C-\frac{1}{2}\Theta_{AC}(Y)A^C.\label{deltaa}
\eea When the vector $Y^A$ is time-independent, the non-local part vanishes. In this case, we find 
the confusing inequality 
\be 
\Delta_A(Y;A;u,\Omega)\not=(\delta_Y-\delta_{f=\frac{1}{2}u\nabla_BY^B})A_A.
\ee 
This is contradictory to the scalar case where \bea 
\Delta(Y;\Sigma;u,\Omega)=(\delta_Y-\delta_{f=\frac{1}{2}u\nabla_BY^B})\Sigma(u,\Omega).
\eea 
Fortunately, the problem is cured by the covariant derivative
\bea  
\Delta_A(Y;A;u,\Omega)=(\delta\hspace{-6pt}\slash_Y-\delta\hspace{-6pt}\slash_{f=\frac{1}{2}u\nabla_BY^B})A_A(u,\Omega).\label{5.44}
\eea  
Therefore, we find the supertranslation and superrotation generators. 
\begin{itemize}
\item  Supertranslation generators $\mathcal{T}_f$. It is the smeared operator of the energy flux density operator $T(u,\Omega)$.
\item Superrotation generators $\mathcal{T}_{\frac{1}{2}u\nabla_AY^A}+\mathcal{M}_Y$. Using the same convention as scalar theory, we will call $\mathcal{M}_Y$ as the superrotation generator. It is a smeared operator of the angular momentum flux density operator $M_A(u,\Omega)$.
\end{itemize} 
We should emphasize that the covariant variation $\delta\hspace{-6pt}\slash$ is necessary for the identification.

\subsection{Symmetry algebra of flux operators}\label{sa}
 It is straightforward to find the following commutators
\bea  
\ [\mathcal{T}_{f_1},\mathcal{T}_{f_2}]&=&C_T(f_1,f_2)+i\mathcal{T}_{f_1\dot f_2-f_2\dot f_1},\label{LL}\\
\ [\mathcal{T}_f,\mathcal{M}_Y]&=&-i\mathcal{T}_{Y^A\nabla_Af}+i\mathcal{M}_{f\dot Y^A}+\frac{i}{2}\mathcal{O}_{\dot Y^A\nabla^Cf\epsilon_{CA}}+\frac{i}{4}\mathcal{Q}_{\frac{d}{du}(\dot Y^A\nabla_A f)},\label{LM}\\
\ [\mathcal{M}_Y,\mathcal{M}_Z]&=&C_M(Y,Z)+i\mathcal{M}_{[Y,Z]}+i\mathcal{O}_{o(Y,Z)}\nn\\&&+\frac{i}{2}\int du du'd\Omega\alpha(u'-u)\Delta_C(\dot Y;A;u',\Omega)\Delta^C(\dot Z;A;u,\Omega).\label{MM}
\eea  
This is not a standard Lie algebra.  There are two new  operators 
\bea  
\mathcal{Q}_h&=&\int du d\Omega h(u,\Omega):A_AA^A:
\eea  and \bea 
\mathcal{O}_g&=&\int du d\Omega g(u,\Omega)\epsilon_{BC}:\dot A^C A^B:\label{Operator1}
\eea  on the right-hand side of the commutators,  where $\epsilon_{BC}$ is the anti-symmetric tensor on the sphere
\begin{align}
    \epsilon_{BC}=\begin{pmatrix}
        0 & \sin\theta \\ -\sin\theta & 0
    \end{pmatrix}.
\end{align}

The first operator $\mathcal{Q}_g$ could compare to the one in the scalar theory. The second operator $\mathcal{O}_g$ is  new. Note that the operators $\mathcal{Q}$ and $
\mathcal{O}$ disappear in \eqref{LM} when $\dot Y=0$. However, the operator $\mathcal{O}$ still appears on the right hand of \eqref{MM} even for $\dot Y=0$. To be more precise, the function 
\bea 
o(Y,Z)=\frac{1}{4}\epsilon^{BC}\Theta_{AB}(Y)\Theta^A_{\ C}(Z)\label{functiono}
\eea is zero only when $Y$ or $Z$ is a CKV. We leave the  discussion  on this new operator in the next section. In this section, we just calculate its commutator with $A_A$
\bea  
\ [\mathcal{O}_g,A_{A'}(u',\Omega')]&=&ig(u',\Omega')\epsilon_{A'B'}A^{B'}(u',\Omega')-\frac{i}{2}\int du \alpha(u-u')\dot{g}(u,\Omega')\epsilon_{A'B'}A^{B'}(u,\Omega')\nn\\
\eea  and the following two-point correlators 
\bea 
\langle T(u,\Omega)O(u',\Omega')\rangle&=&0,\label{TO2pt}\\
\langle M_A(u,\Omega)O(u',\Omega')\rangle&=&\frac{\beta(u-u')-\frac{1}{4\pi}}{4\pi(u-u'-i\epsilon)^2}\epsilon_{AC}\nabla^C\delta(\Omega-\Omega')\delta^{(2)}(0),\\
\langle O(u,\Omega)O(u',\Omega')\rangle&=&-\frac{\beta(u-u')-\frac{1}{4\pi}}{2\pi(u-u'-i\epsilon)^2}\delta(\Omega-\Omega')\delta^{(2)}(0), \label{OO2pt}
\eea where the operator $O(u,\Omega)=\epsilon_{AB}:\dot A^BA^A:$.
Now it is straightforward to find\footnote{Details can be found in Appendix \ref{comsec}.} 
\bea  
\ [\mathcal{T}_{f_1},\mathcal{T}_{f_2}]&=&C_T(f_1,f_2)+i\mathcal{T}_{f_1\dot f_2-f_2\dot f_1},\label{LL2}\\
\ [\mathcal{T}_f,\mathcal{M}_Y]&=&-i\mathcal{T}_{Y^A\nabla_Af}+i\mathcal{M}_{f\dot Y^A}+\frac{i}{2}\mathcal{O}_{\dot Y^A\nabla^Cf\epsilon_{CA}}+\frac{i}{4}\mathcal{Q}_{\frac{d}{du}(\dot Y^A\nabla_A f)},\label{LM2}\\
\ [\mathcal{T}_f,\mathcal{O}_g]&=&i\mathcal{O}_{f\dot g},\label{LO}\\
\ [\mathcal{M}_Y,\mathcal{M}_Z]&=&C_M(Y,Z)+i\mathcal{M}_{[Y,Z]}+i\mathcal{O}_{o(Y,Z)}\nn\\&&+\frac{i}{2}\int du du'd\Omega\alpha(u'-u)\Delta_C(\dot Y;A;u',\Omega)\Delta^C(\dot Z;A;u,\Omega),\label{MM2}\\
\ [\mathcal{M}_Y,\mathcal{O}_g]&=&C_{MO}(Y,g)+i\mathcal{O}_{Y^A\nabla_A g}\nn\\&&-\frac{i}{2}\int du du'd\Omega \alpha(u'-u)\dot g(u,\Omega)\Delta^B(\dot Y;A;u',\Omega)\epsilon_{BC}A^C(u,\Omega),\\
\ [\mathcal{O}_{g_{{}_1}},\mathcal{O}_{g_{{}_2}}]&=&C_O(g_1,g_2)+\frac{i}{2}\int du du'd\Omega\alpha(u'-u)\dot{g}_2(u,\Omega)\dot{g}_1(u',\Omega)A_C(u,\Omega)A^C(u',\Omega),\label{OO}
\eea  
We will discuss the commutators in the following.
\begin{itemize}
\item Central charges. The central extension terms can be derived from the two-point functions \eqref{TT2pt}-\eqref{MM2pt} and \eqref{TO2pt}-\eqref{OO2pt}. 
There are four non-vanishing central extension terms \bea  
C_T(f_1,f_2)&=&-\frac{i}{24\pi}c\ \mathcal{I}_{f_1{\dddot f}_{\hspace{-4pt}2}-f_2{\dddot f}_{\hspace{-4pt}1}},\label{CLf1f2}\\
C_O(g_1,g_2)&=&4c\int du du'd\Omega g_1(u,\Omega)g_2(u',\Omega)\eta(u-u'),\\
C_M(Y,Z)&=&2\int du du'd\Omega d\Omega' Y^A(u,\Omega)Z^{B'}(u',\Omega')\kappa_{AB'}(\Omega,\Omega')\eta(u-u'),\\
C_{MO}(Y,g)&=&-2c\int du du'd\Omega Y^A(u,\Omega)\nabla^C g(u',\Omega)\epsilon_{AC}\eta(u-u') 
\eea where the function 
\be  
\eta(u-u')=-\frac{\beta(u-u')-\frac{1}{4\pi}}{8\pi(u-u'-i\epsilon)^2}+\frac{\beta(u'-u)-\frac{1}{4\pi}}{8\pi(u'-u-i\epsilon)^2}
\ee  and the identity operator 
\be 
\mathcal{I}_f=\int du d\Omega f(u,\Omega)
\ee have already been defined in the scalar theory.  
We use a constant $c$ to denote the divergent part 
\be 
c=\delta^{(0)}(0).
\ee 

\item Virasoro algebra. By transforming to the Fourier space, the equation \eqref{LL2} implies a higher dimensional Virasoro algebra 
\bea \  [\mathcal{T}_{\omega,\ell,m},\mathcal{T}_{\omega',\ell',m'}]=(\omega'-\omega)\sum_{L=|\ell-\ell'|}^{\ell+\ell'}\sum_{M=-L}^L c_{\ell,m;\ell',m';L,M}\mathcal{T}_{\omega+\omega',L,M}-(-1)^m\frac{\omega^3}{6}c\ \delta(\omega+\omega')\delta_{\ell,\ell'}\delta_{m,-m'}.\nn\\
\eea 
The constants $c_{\ell,m;\ell',m';L,M}$ are Clebsch-Gordan coefficients. 
There are two propagating degrees of freedom in the vector theory, hence the central term is two times compared to the real scalar theory.
 
\item Non-local terms. There are three non-local terms in \eqref{MM2}-\eqref{OO}. The non-local terms introduce  new operators in the commutators. It is understood that the new operators are also normal ordered. Interestingly, the non-local term in \eqref{MM2} has the same structure as scalar theory. It would be interesting to explore the physical origin of this fact. There is also an interesting truncation by setting 
\be 
\dot Y=\dot Z=\dot g_1=\dot{g}_2=0.\label{nnon}
\ee In this case, all the non-local terms and the central terms $C_M,C_O,C_{MO}$ are vanishing. The reader can find more details in \cite{Liu:2022mne}.
\end{itemize}

\subsection{Closed algebra}
Due to the existence of non-local terms and the physically meaningless operator $\mathcal{Q}_h$, the aforementioned algebra \eqref{LL2}-\eqref{OO} is not closed. As we have shown, requiring \eqref{nnon} will make these annoying terms vanish, and hence lead to a closed algebra.

\paragraph*{Truncation \Rmnum{1}.} 
By imposing the condition \eqref{nnon}, we find the following truncated algebra  
\bea 
\    [\mathcal{T}_{f_1},\mathcal{T}_{f_2}]&=&C_T(f_1,f_2)+i\mathcal{T}_{f_1\dot f_2-f_2\dot f_1},\label{vir1}\\
\ [\mathcal{T}_f,\mathcal{M}_Y]&=&-i\mathcal{T}_{Y^A\nabla_Af},\label{vir2}\\
\ [\mathcal{M}_Y,\mathcal{M}_Z]&=&i\mathcal{M}_{[Y,Z]}+i\mathcal{O}_{o(Y,Z)},\label{vir3}\\
\ [\mathcal{T}_f,\mathcal{O}_g]&=&0,\\
\ [\mathcal{M}_Y,\mathcal{O}_g]&=&i\mathcal{O}_{Y^A\nabla_A g},\label{575}\\
\ [\mathcal{O}_{g_1},\mathcal{O}_{g_2}]&=&0.\label{viroo}
\eea This is an enlarged algebra compared to the one found in the scalar theory. The Jacobi identities are checked  in Appendix \ref{smvsec}. In the scalar theory, the operator $\mathcal{T}_f$ generates GSTs and $\mathcal{M}_Y$ generates SSRs. The corresponding group  is 
\be 
\text{Diff}(S^2)\ltimes C^\infty(\mathcal{I}^+) \label{virnu}
\ee  where the notation $\text{Diff}(S^2)$ means that the vectors $Y^A(\Omega)$ generate diffeomorphisms of $S^2$ and $C^\infty(\mathcal{I}^+)$ means that $f$ is any smooth function on $\mathcal{I}^+$. In the vector theory, although the term $\frac{i}{2}\mathcal{O}_{\dot Y^A\nabla^Cf\epsilon_{CA}}$ in \eqref{LM2} now vanishes, the appearance of operator $\mathcal{O}$ on the right-hand side of \eqref{vir3} indicates that the enhancement of the group \eqref{virnu} is unavoidable. 

One may be interested in this new operator. As we will show in the next section, this operator can be derived from the electromagnetic duality transformations. It is amazing that the commutator of superrotations will produce a term reflecting internal symmetry. However, we must point out that the superrotation flux operators do not agree with ordinary variation when acting on vector fields. As is shown in \eqref{myaa}, for $\dot Y=0$, we have
\begin{align}
    [\mathcal{M}_Y,A_{A'}(u',\Omega')]=-i\Delta_{A'}(Y;A;u',\Omega'),
\end{align}
where $\Delta_{A'}(Y;A;u',\Omega')$ is given by \eqref{deltaa}. Note that it is not the one induced by Lie derivative. Namely, the superrotation variation needs to be corrected according to the principle of covariant variation. 
Besides,  $\mathcal{O}_g$ generates original EM duality transformations when $g$ is a constant. In this case, the right-hand side of \eqref{575} vanishes. It makes sense since original EM duality is expected to have nothing to do with geometric transformations. 

\paragraph*{Truncation \Rmnum{2}.} To eliminate the operator $\mathcal{O}$, we should require 
\be 
o(Y,Z)=0.
\ee This implies that $Y, Z$ are CKVs.  In this case, the truncated algebra becomes 
\bea 
\    [\mathcal{T}_{f_1},\mathcal{T}_{f_2}]&=&C_T(f_1,f_2)+i\mathcal{T}_{f_1\dot f_2-f_2\dot f_1},\label{vir1ckv}\\
\ [\mathcal{T}_f,\mathcal{M}_Y]&=&-i\mathcal{T}_{Y^A\nabla_Af},\label{vir2ckv}\\
\ [\mathcal{M}_Y,\mathcal{M}_Z]&=&i\mathcal{M}_{[Y,Z]},\label{vir3ckv}
\eea  
This is the Newmann-Unti group with a central charge
\be 
\text{NU}(\mathcal{I}^+,\gamma,\chi)=\text{Conf}(S^2)\ltimes C^\infty(\mathcal{I}^+).
\ee 
As one expects, we obtain a geometric algebra. It is consistent with the fact that the connection \eqref{connection} vanishes for $Y$ being a CKV. In other words, for Lorentz transformations, the covariant variation agrees with the one induced by Lie derivative, and there are no additional terms in the algebra.

\paragraph*{Other truncations.} One may further truncate the aforementioned Newmann-Unti group to the one of level $k$ as has been done in \cite{Liu:2022mne}. Besides, one can also demand $\dot f=0$ and $Y,Z$ to be CKVs, which leads to the BMS algebra.

\section{Electromagnetic duality operator}\label{emdualsec}
The electromagnetic duality (EM duality) is a symmetry transformation for sourceless Maxwell equation \cite{Oliver:1892,Dirac:1931kp,Dirac:1948um,1957AnPhy...2..525M,Deser:1976iy}. Duality invariance of Maxwell equation leads to the introduction of magnetic monopole and the quantization of electric charge \cite{Dirac:1931mon}. It has been elaborated in  non-Abelian gauge theories by \cite{Deser:1976iy}.  One can find more details in \cite{Olive:1995sw}. 

EM duality  transformation exchanges the role of the  electric and magnetic field 
\bea  
e_i\to e'_i=\cos\varphi e_i+ \sin\varphi b_i,\quad  b_i\to  b'_i=-\sin\varphi e_i+\cos\varphi  b_i \label{dual}
\eea  where $\varphi$ is a constant. The $SO(2)$ rotation \eqref{dual} is equivalent to the following phase transformation 
\bea  
f_i\to f_i'=e^{-i\varphi}f_i,\quad \bar{f}_i\to \bar{f}_i'= e^{i\varphi}\bar{f}_i,
\eea  where 
\be  
f_i=e_i+ib_i,\quad \bar{f}_i=e_i-ib_i.
\ee  At $\mathcal{I}^+$, this reduces to 
\bea  
F_i\to e^{-i\varphi}F_i,\quad \bar{F}_i\to e^{i\varphi}\bar{F}_i\label{duality0}
\eea  with
\bea  
F_i=E_i+iB_i&=&(Y_i^A-i\widetilde{Y}_i^A)\dot A_A,\\
\bar{F}_i=E_i-iB_i&=&(Y_i^A+i\widetilde{Y}_i^A)\dot A_A.
\eea    Since the duality transformation \eqref{dual} is a continuous global symmetry, there may be a corresponding conserved current by Noether's theorem. However, it is easy to check that the original action 
\be 
S[a]=-\frac{1}{4}\int d^4x f_{\mu\nu}f^{\mu\nu}\label{originalaction}
\ee is not invariant under the EM duality transformation. To find the corresponding current,  we may introduce a dual EM vector field $\widetilde{a}_\mu$ and its corresponding EM field $\widetilde{f}_{\mu\nu}$
\be \widetilde{f}_{\mu\nu}=\partial_\mu\widetilde{a}_\nu-\partial_\nu\widetilde{a}_\mu.\ee 
The dual field $\widetilde{f}_{\mu\nu}$ is invariant under the dual gauge transformation 
\be 
\delta_{\widetilde{\epsilon}}\widetilde{a}_\mu=\partial_\mu\widetilde{\epsilon}.\label{dualgauge}
\ee  Following \cite{Bliokh:2012zr}, we write a symmetric action for (sourceless) electromagnetic theory  
\bea 
S[a,\widetilde{a}]=-\frac{1}{8}\int d^4x (f_{\mu\nu}f^{\mu\nu}+\widetilde{f}_{\mu\nu}\widetilde{f}^{\mu\nu}).\label{dualaction}
\eea  Treating the vector fields $a_\mu,\widetilde{a}_\mu$ as independent quantities, 
the equation of motions from the symmetric action \eqref{dualaction} become
\bea 
\partial_\mu f^{\mu\nu}=\partial_\mu\widetilde{f}^{\mu\nu}=0,\quad \partial_{[\mu}f_{\nu\rho]}=\partial_{[\mu}\widetilde{f}_{\nu\rho]}=0.\label{dualequation}
\eea They are equivalent to the sourceless Maxwell equations
 with an additional constraint 
\be  
\widetilde{f}_{\mu\nu}=-\frac{1}{2}\epsilon_{\mu\nu\rho\sigma}f^{\rho\sigma}.\label{dualitymatch}
\ee 
 The constraint relates the field $\widetilde{f}_{\mu\nu}$  to the Hodge dual of the field $f_{\mu\nu}$.
Note that
the symmetric action is not equal to the original action \eqref{originalaction} when the constraint is imposed. Nevertheless, it is invariant under EM duality transformation and turns out to be useful to derive the corresponding  conserved currents. The EM duality transformation may be expressed as
\be  
a_\mu\to a'_\mu=\cos\varphi a_\mu+\sin\varphi\widetilde{a}_\mu,\quad \widetilde{a}_\mu\to \widetilde{a}_\mu'=-\sin\varphi a_\mu+\cos\varphi \widetilde{a}_\mu.
\ee 
 The conserved current for the EM duality could be found in \cite{2005math.ph...1052A,Bliokh:2012zr} 
 by using  Noether's theorem 
 \bea  
j_{\text{em}}^\mu=\frac{1}{2}(f^{\mu\nu}\widetilde{a}_\nu-\widetilde{f}^{\mu\nu}a_\nu).\label{dualitycurrent}
\eea
 The conserved charge corresponding to the current 
\be 
\int (d^3x)_\mu\ j^\mu_{\text{em}}
\ee is called optical helicity. At the microscopic level, this is the difference between the number of the photons with left helicity and right helicity.
To obtain the relation between vector field and its dual, we may first use the dual gauge transformation \eqref{dualgauge} to fix 
\be  
\widetilde{a}_r=0.
\ee  In retarded coordinates, we impose the fall-off condition for the dual field
\bea  
\widetilde{a}_u&=&\frac{\widetilde{A}_u(u,\Omega)}{r}+\mathcal{O}\left(\frac{1}{r^2}\right),\\
\widetilde{a}_A&=&\widetilde{A}_A(u,\Omega)+\mathcal{O}\left(\frac{1}{r}\right).
\eea 
To satisfy the constraint condition \eqref{dualitymatch}, we should identify 
\bea  
\widetilde{A}_A=A^C\epsilon_{CA}.
\eea This is a Hodge dual on the unit sphere. Now we may use the conservation of the EM duality current 
\be 
\partial_\mu j_{\text{em}}^\mu=0
\ee 
 to find the EM duality fluxes which is radiated to  $\mathcal{I}^+$
\bea  
\int d\Omega r^2 n_i j_{\text{em}}^i&=&\int d\Omega \dot A^C A^B\epsilon_{CB}.\label{dualityflux}
\eea  
Obviously, this looks like the second new operator \eqref{Operator1} and that is why we discuss the EM duality transformation. Actually, the EM duality transformation and its generator have been studied from other starting points, and we refer readers to \cite{Hosseinzadeh:2018dkh,Henneaux:2020nxi,Maleknejad:2023nyh,Oblak:2023axy}.

Now we define the EM duality flux density operator 
\be
O(u,\Omega)=:\dot A^C A^B\epsilon_{BC}:\label{Operator}
\ee and use it to construct the smeared operator 
\be 
\mathcal{O}_g=\int du d\Omega g(u,\Omega) O(u,\Omega).
\ee
The commutators between $\mathcal{O}_g$ and $F_i,\bar{F}_i$ are 
\bea  
\ [\mathcal{O}_g,F_i]&=&-g F_i-\frac{1}{2}\dot g G_i,\label{duality1}\\
\ [\mathcal{O}_g,\bar{F}_i]&=&g \bar{F}_i+\frac{1}{2}\dot g\bar{G}_i.\label{duality2}
\eea   where 
\bea \dot G_i=F_i,\quad \dot{\bar{G}}_i=\bar{F}_i.
\eea 
\begin{itemize}
\item 
When $g$ is a constant, the transformation \eqref{duality1}-\eqref{duality2} is  exactly the infinitesimal EM duality transformation \eqref{duality0}.
\item When $g$ is a time-independent function 
\be 
g=g(\Omega),
\ee the transformation \eqref{duality1}-\eqref{duality2} would be angle-dependent. This is a generalized EM duality transformation at $\mathcal{I}^+$. 
\item When $g$ is time-dependent, the additional terms in \eqref{duality1}-\eqref{duality2} obscure the interpretation of the operator.  
\end{itemize}

\section{Antipodal matching condition}\label{antipodalsec}

The symmetry group can also be discussed at past null infinity $(\mathcal{I}^-)$. The fall-off conditions \eqref{falloffdicar} can also be expressed near $\mathcal{I}^-$ 
\bea 
a_\mu(v,r,\Omega)&=&\frac{A^-_\mu(v,\Omega)}{r}+\sum_{k=2}^\infty \frac{A_\mu^{-(k)}(v,\Omega)}{r^k},\quad \mu=0,1,2,3,
\eea where $v=t+r$ is the advanced time. The supertranslation and superrotation generators depend on the first two leading orders of the vector potential.  We may use the large-$r$ expansion of the spherical Bessel function of the first kind
\bea 
j_\ell(k r)=\frac{\sin(kr-\frac{\pi\ell}{2})}{kr}+\frac{\ell(\ell+1)}{2k^2r^2}\cos(kr-\frac{\pi\ell}{2})+\mathcal{O}\left(\frac{1}{r^3}\right)\label{expansionsphericalbessel}
\eea  
and the mode expansion \eqref{modeexpansion} to find
\bea 
A_\mu(u,\Omega)&=&\int_0^\infty \frac{d\omega}{\sqrt{4\pi\omega}} \sum_{\ell,m}[c_{\mu;\omega,\ell,m}e^{-i\omega u}Y_{\ell,m}(\Omega)+\text{h.c.}],\\
A_\mu^{-}(v,\Omega)&=&\int_0^\infty \frac{d\omega}{\sqrt{4\pi\omega}} \sum_{\ell,m}[\widetilde{c}_{\mu;\omega,\ell,m}e^{-i\omega v}Y_{\ell,m}(\Omega)+\text{h.c.}] ,\\
A_\mu^{(2)}(u,\Omega)&=&\int_0^\infty \frac{d\omega}{\sqrt{4\pi\omega}}\sum_{\ell,m}[\frac{i\ell(\ell+1)}{2\omega}c_{\mu;\omega,\ell,m}e^{-i\omega u}Y_{\ell,m}(\Omega)+\text{h.c.}],\\
A_\mu^{-(2)}(v,\Omega)&=&\int_0^\infty \frac{d\omega}{\sqrt{4\pi\omega}}\sum_{\ell,m}[\frac{\ell(\ell+1)}{2i\omega}\widetilde{c}_{\mu;\omega,\ell,m}e^{-i\omega v}Y_{\ell,m}(\Omega)+\text{h.c.}]
\eea 
with 
\bea 
c_{\mu;\omega,\ell,m}&=&\frac{\omega}{(2\pi)^{3/2}i}\int d\Omega \sum_{\alpha=\pm}\epsilon_\mu^{*\alpha}(\bm k)b_{\alpha,\bm k}Y_{\ell,m}^*(\Omega),\\
c_{\mu;\omega,\ell,m}^\dagger&=&\frac{i\omega}{(2\pi)^{3/2}}\int d\Omega \sum_{\alpha=\pm}\epsilon_\mu^{\alpha}(\bm k)b^\dagger_{\alpha,\bm k}Y_{\ell,m}(\Omega),\\
\widetilde{c}_{\mu;\omega,\ell,m}&=&(-1)^\ell\frac{i\omega}{(2\pi)^{3/2}}\int d\Omega \sum_{\alpha=\pm}\epsilon_\mu^{*\alpha}(\bm k)b_{\alpha,\bm k}Y_{\ell,m}^*(\Omega),\\
\widetilde{c}_{\mu;\omega,\ell,m}^\dagger&=&(-1)^\ell \frac{\omega}{(2\pi)^{3/2}i}\int d\Omega \sum_{\alpha=\pm}\epsilon_\mu^{\alpha}(\bm k)b^\dagger_{\alpha,\bm k}Y_{\ell,m}(\Omega).
\eea The creation and annihilation operators are related by 
\be 
c_{\mu;\omega,\ell,m}=(-1)^{\ell+1}\widetilde{c}_{\mu;\omega,\ell,m}.
\ee In frequency space, we have
\bea 
A_\mu(\omega,\Omega)=-A_\mu^-(\omega,\Omega^P),\quad A_\mu^{(2)}(\omega,\Omega)=A_\mu^{-(2)}(\omega,\Omega^P)\label{freAmu}
\eea 
where $\Omega^P$ is the antipodal point of $\Omega$
\be 
\Omega^P=(\pi-\theta,\pi+\phi).
\ee
The electric and magnetic fields near $\mathcal{I}^-$ are expanded as 
\bea 
e_i&=&-f_{0i}\equiv \frac{E_i^{-}(v,\Omega)}{r}+\sum_{k=2}^\infty \frac{E_i^{-(k)}(v,\Omega)}{r^k},\\
b_i&=&\frac{1}{2}\epsilon_{ijk}f^{jk}=\frac{B_i^-(v,\Omega)}{r}++\sum_{k=2}^\infty \frac{B_i^{-(k)}(v,\Omega)}{r^k},
\eea where \bea 
E_i^-(v,\Omega)&=&n_i\dot A^-_0-\dot A^-_i,\label{EiA0m}\\
E_i^{-(2)}(v,\Omega)&=&-Y_i^A\partial_AA^-_0(v,\Omega)-n_i A^-_0(v,\Omega)+n_i \dot{A}_0^{-(2)}(v,\Omega)-\dot{A}_i^{-(2)}(v,\Omega),\\
B_i^{-}(v,\Omega)&=&\epsilon_{ijk}n_j\dot{A}_k^-(v,\Omega),\\
B_i^{-(2)}(v,\Omega)&=&-\epsilon_{ijk}Y_j^A\partial_A A^-_k-\epsilon_{ijk}n_j A^-_k+\epsilon_{ijk}n_j\dot{A}_k^{-(2)}.\label{Bi2Aim}
\eea 
In Fourier space, the electric and magnetic fields \eqref{EiA0}-\eqref{Bi2Ai} and \eqref{EiA0m}-\eqref{Bi2Aim} are
\bea 
E_i(\omega,\Omega)&=&i\omega [n_i(\Omega)A_0(\omega,\Omega)+A_i(\omega,\Omega)],\\
E_i^{(2)}(\omega,\Omega)&=&-Y_i^A(\Omega)\partial_A A_0(\omega,\Omega)-n_i(\Omega)A_0(\omega,\Omega)+i\omega n_i(\Omega)A_0^{(2)}(\omega,\Omega)+i\omega A_i^{(2)}(\omega,\Omega),\nn\\
B_i(\omega,\Omega)&=&i\omega\epsilon_{ijk}n_j(\Omega)A_k(\omega,\Omega),\\
B_i^{(2)}(\omega,\Omega)&=&-\epsilon_{ijk}Y_j^A(\Omega)\partial_AA_k(\omega,\Omega)-\epsilon_{ijk}n^j(\Omega)A_k(\omega,\Omega)+i\omega \epsilon_{ijk}n_j(\Omega)A_k^{(2)}(\omega,\Omega),\nn\\
E_i^-(\omega,\Omega)&=&i\omega [-n_i(\Omega)A^-_0(\omega,\Omega)+A^-_i(\omega,\Omega)],\\
E_i^{-(2)}(\omega,\Omega)&=&-Y_i^A(\Omega)\partial_AA_0^-(\omega,\Omega)-n_i(\Omega)A_0^-(\omega,\Omega)-i\omega n_i(\Omega)A_0^{-(2)}(\omega,\Omega)+i\omega A_i^{-(2)}(\omega,\Omega),\nn\\
B_i^-(\omega,\Omega)&=&-i\omega \epsilon_{ijk}n_j(\Omega)A_k^-(\omega,\Omega),\\
B_i^{-(2)}(\omega,\Omega)&=&-\epsilon_{ijk}Y_j^A(\Omega)\partial_AA_k^-(\omega,\Omega)-\epsilon_{ijk}n_j(\Omega)A^-_k(\omega,\Omega)-i\omega \epsilon_{ijk}n_j(\Omega)A_k^{-(2)}(\omega,\Omega).\nn
\eea 
Using the relations
\bea 
n_i(\Omega^P)=-n_i(\Omega),\quad Y^A_i(\Omega)\partial_A=Y^A_i(\Omega^P)\partial_A^P
\eea and the matching condition \eqref{freAmu},
the antipodal matching condition for the electric and magnetic fields is
\bea 
E_i(\omega,\Omega)&=&-E_i^-(\omega,\Omega^P),\\
B_i(\omega,\Omega)&=&-B_i^-(\omega,\Omega^P),\\
E_i^{(2)}(\omega,\Omega)&=&E_i^{-(2)}(\omega,\Omega^P),\\
B_i^{(2)}(\omega,\Omega)&=&B_i^{-(2)}(\omega,\Omega^P).
\eea The antipodal matching condition can also be checked using Green's functions. One can find the details in Appendix \ref{greensec}.
\section{Conclusion and discussion}\label{dissec}
In this paper, we reduce the electromagnetic field theory in Minkowski spacetime to future null infinity $\mathcal{I}^+$.  The boundary vector theory is characterized by a single vector field $A_A$ with a non-trivial symplectic form.  The ten Poincar\'e fluxes are totally determined by the field $A_A$. We obtain the flux operators and interpret them as supertranslation and superrotation generators. Interestingly, one should define a covariant variation to identify the superrotation generators. The supertranslation and superrotation flux operators do not form a closed algebra for GSTs and GSRs.  In contrast to the scalar field theory, the  GSTs and SSRs cannot form a closed group. One should introduce a new operator which corresponds to a generalized EM duality transformation at $\mathcal{I}^+$. By combining the GSTs, SSRs as well as the generalized EM duality transformations, we could find a closed group whose Lie algebra has been given in \eqref{vir1}-\eqref{viroo}. 

There is a no-go theorem which is presented in \cite{Schwarz:2022dqf} recently. It states that the conformal symmetry of the holographic theory on the celestial sphere must not be extended to diffeomorphism symmetry since the $\text{Diff}(S^2)$ implies the conservation of the conformal spin. Our work bypasses the no-go theorem in two ways. At first, the vector theory that we find is free at $\mathcal{I}^+$. Secondly, the diffeomorphism is   intertwined with the EM duality transformation for the boundary vector theory. The EM duality symmetry is broken in the interacting theory. Therefore, the symmetry group we find in this work may break when there is interaction. It would be interesting to explore the interacting vector theory in the future.

There are various open questions in this direction. 
\begin{itemize}
	\item Covariant variation. The introduction of the covariant variation $\delta\hspace{-6pt}\slash$ is rather interesting. There is a natural variation $\delta$  at $\mathcal{I}^+$ which is induced by the diffeomorphism from the bulk.
 The variation $\delta$ has a direct geometric meaning. However, it is not always metric compatible. 
 To cure this problem, we define a connection $\Gamma_{AB}$ such that the covariant variation is 
 \be 
 \delta\hspace{-6pt}\slash=\delta+\text{connections}
 \ee schematically. This is quite similar to the definition of the covariant derivative $\nabla$ in general relativity
 \bea 
 \nabla=\partial+\text{connections}.
 \eea We note the connection term $\Gamma_{AB}$ is proportional to the variation of the metric $\gamma$ and it is non-zero when the superrotation vector is not a CKV. The consequence of the covariant variation is that the commutator between two superrotation generators is not a superrotation generator. Equation \eqref{vir3} may be read schematically as 
 \bea 
 \ [\text{superrotation},\text{superrotation}]=\text{superrotation}+\text{generalized EM duality}.
 \eea 
 Using the notation $\Delta_Y=\delta\hspace{-6pt}\slash_Y-\delta\hspace{-6pt}\slash_{f=\frac{1}{2}u\nabla\cdot Y}$, this result may be rewritten as 
 \bea 
\ ([\Delta_Y,\Delta_Z]-\Delta_{[Y,Z]})(\cdots)=R(Y,Z)(\cdots)
 \eea 
We have introduced a  formal curvature tensor $R(Y,Z)$ similar to the case of covariant derivative. It is rather interesting to understand why the generalized EM duality operator is related to the curvature tensor.

	\item Field theories on the Carroll manifold $\mathcal{I}^+$. The field theory on $\mathcal{I}^+$ may provide an explicit realization of flat holography. Carrollian diffeomorphism has a direct geometric meaning which is enough for constructing scalar theory. Our result implies that Carrollian diffeomorphism $\text{Diff}(S^2)\ltimes {C}^\infty(\mathcal{I}^+)$ is not the end of the story for theories with non-zero spin. The most intriguing case would be to project the gravitational theory to its boundary. We will present the result in the near future. 
 \item  Large gauge transformation. Besides the diffeomorphism, the electromagnetic theory is also invariant under $U(1)$ gauge transformation. The gauge invariance is broken at $\mathcal{I}^+$ and part of the gauge transformations become large gauge transformations. As a consequence, there is an infinite-dimensional algebra  \cite{He:2014cra,Kapec:2014zla,Kapec:2015ena,Campiglia:2015qka,Kapec:2017tkm,Nande:2017dba,Henneaux:2018gfi,Freidel:2018fsk,Fuentealba:2023rvf} at the boundary. Our work shows that there is also an extended algebra coming from diffeomorphism. Therefore, it   may be natural to combine the two results \cite{Ferko:2021bym} in the future. 

 \item 
Divergences. There are two kinds of divergences appearing in the context. The first one is about the correlation function of two fields, i.e. $\braket{0|A_AA_B|0}$. It is divergent due to the appearance of $\beta(u-u')$. Taking time derivative will eliminate this divergence, and the same is true for taking the difference, as we have shown in \eqref{alphabeta}. Actually, we could deal with the divergence of $\beta(u-u')$ in two different ways. The first one is similar to dimensional regularization, i.e., adding an infinitesimal parameter $\kappa$ such that to order $\mathcal{O}(\kappa^0)$, we have
\begin{equation}
  \begin{aligned}
    \beta(u-u')=&\lim_{\kappa\to 0}\int_0^\infty \frac{d\omega}{4\pi\omega^{1-\kappa}}e^{-i\omega(u-u'-i\epsilon)}\\=&\frac{1}{4\pi\kappa}-\frac{1}{4\pi}\log(i(u-u'-i\epsilon))-\frac{\gamma_E}{4\pi},
  \end{aligned}
\end{equation}
with $\gamma_E$ denoting Euler constant. The divergent term $\frac{1}{4\pi\kappa}$ and constant terms may be absorbed into a constant $-\frac{1}{4\pi}\log\omega_0$, and we find a finite result
\bea 
\beta(u-u')=-\frac{1}{4\pi}\log (\omega_0(u-u'-i\epsilon)).
\eea 
The second way is to introduce an infrared cutoff $\omega'_0\to0$ for the integral which is similar to the Pauli-Villars regularization. This makes sense since the divergence comes from integration in the region of little $\omega$.
It follows that to order $\mathcal{O}(1)$, we have
\begin{equation}
  \begin{aligned}
    \beta(u-u')=&\lim_{\omega'_0\to 0^+}\int_{\omega'_0}^\infty \frac{d\omega}{4\pi\omega} e^{-i\omega(u-u'-i\epsilon)}\\
    =&-\frac{1}{4\pi}\log \omega'_0-\frac{1}{4\pi}\log(i(u-u'-i\epsilon))-\frac{\gamma_E}{4\pi}\nn\\
    =&-\frac{1}{4\pi}\log (\omega_0(u-u'-i\epsilon))
  \end{aligned}
\end{equation}
We have absorbed the first term and constant terms to a constant $-\frac{1}{4\pi}\log\omega_0$ again. These two  ways lead to the same result, and they both agree with the fact that the time derivatives or the difference of $\beta(u-u')$ are finite.

Secondly, there is a divergent factor $c=\delta^{(2)}(0)$ in some central charges, which comes from two Dirac delta functions in the angular direction, appearing in the four-point correlators. As we have analyzed in the scalar theory \cite{Liu:2022mne}, this function can be obtained from the summation of spherical functions with the same arguments, i.e., $\delta^{(2)}(0)=\sum_{\ell,m}Y_{\ell,m}(\Omega)Y_{\ell,m}^*(\Omega)$. Making use of the addition theorem, we find
\begin{align}
    \delta^{(2)}(0)=\frac{1}{4\pi}\sum_{\ell=0}^\infty (2\ell+1)=\frac{1}{4\pi}\sum_{\ell,m} 1.
\end{align}
The  denominator $4\pi$ equals the area of a unit sphere, and thus $\delta^{(2)}(0)$ can be interpreted as the state density on unit sphere. Moreover, as a naive method, one may use Riemann zeta function to regularize $\delta^{(2)}(0)$. From the classic evaluations $\zeta(-1)=-1/12,\ \zeta(0)=-1/2$, one gets a finite value $\delta^{(2)}(0)=\frac{1}{12\pi}$. 

\end{itemize}

\vspace{10pt}
{\noindent \bf Acknowledgments.} 
The work of J.L. is supported by NSFC Grant No. 12005069.
\appendix
\section{Vector fields on \texorpdfstring{$S^2$}{}}
\label{vfs}
The metric for a unit $S^2$ is 
\be 
ds_{S^2}^2=\gamma_{AB}d\theta^Ad\theta^B,\quad A,B=1,2.
\ee It can be embedded into the Euclidean space $\mathbb{R}^3$
\be 
ds^2_{\mathbb{R}^3}=\delta_{ij}dx^idx^j,\quad i,j=1,2,3
\ee 
by the map
\bea 
x^i=n^i,
\eea  where $n^i$ is the unit normal vector of $S^2$.

\subsection{Conformal Killing vectors}
\label{ckvs}
A CKV is the vector $Y^A$ that obeys the equation 
\bea 
\Theta_{AB}(Y)=0.
\eea There are six global solutions for this equation. 
\begin{itemize}
	\item There are three Killing vectors on $S^2$
 \bea 
Y_{12}^A&=&(0,1),\\
Y^A_{23}&=&(-\sin\phi,-\cot\theta\cos\phi),\\
Y^A_{13}&=&(-\cos\phi,\cot\theta\sin\phi).
\eea They are denoted as $Y^A_{ij}$ in the context. The subscript $ij$ are antisymmetric\bea 
Y^A_{ij}=-Y^A_{ji},\quad i,j=1,2,3.
\eea 
They satisfy the condition 
\be 
\nabla_A Y^A_{ij}=0.
\ee We can also use the Levi-Civita tensor $\epsilon_{ijk}$ to construct the equivalent Killing vectors
\bea 
\widetilde{Y}^A_i=\frac{1}{2}\epsilon_{ijk}Y_{jk}^A.
\eea 
 \item There are also three strictly conformal Killing vectors on $S^2$
 \bea 
Y_1^A&=&(-\cos\theta\cos\phi,\frac{\sin\phi}{\sin\theta}),\\Y_2^A&=&(-\cos\theta\sin\phi,-\frac{\cos\phi}{\sin\theta}),\\ Y_3^A&=&(\sin\theta,0).
\eea
 They are denoted as $Y^A_i$ in the context. The subscript $i=1,2,3$. Their divergences are non-zero
 \be 
 \nabla_A Y_i^A=2n_i.
 \ee 
\end{itemize} 
We list  the related properties in the following. 
\begin{enumerate}
\item Commutators. 
The six CKVs {form} a closed lie algebra which is isomorphic to $so(1,3)$
\bea 
\ [Y_i,Y_j]&=&Y_{ij},\nn\\ \ [Y_{ij},Y_k]&=&-\delta_{ik}Y_j+\delta_{jk}Y_i,\\ \ [Y_{ij},Y_{kl}]&=& -\delta_{ik}Y_{jl}+\delta_{jk}Y_{il}-\delta_{jl}Y_{ik}+\delta_{il}Y_{jk}.
\eea 
\item  Relations. The three Killing vectors and the three strictly CKVs are related to each other by the identities 
\bea 
Y_i^A=Y_{ij}^A n_j,\quad Y_{ij}^A=Y_i^An_j-Y_j^An_i.
\eea 
The vector $Y_i^A,\widetilde{Y}_i^A$ are related by  the Levi-Civita tensor 
\bea  
\widetilde{Y}_{iC}=Y_{i}^A\epsilon_{AC},\quad Y_{iC}=-\widetilde{Y}_i^A\epsilon_{AC}.
\eea 
From these, it is easy to find
\begin{align}
    Y^A_i\widetilde{Y}^B_i=0.
\end{align}

\item 
The six CKVs are related to the metric $\gamma^{AB}$ and $\delta_{ij}$ by 
\bea 
Y_i^A Y_i^B=\gamma^{AB},\quad Y_i^AY_j^B\gamma_{AB}+n_in_j=\delta_{ij}.
\eea 
\item The vector $Y_i^A$ is orthogonal to the normal vector $n_i$ 
\be  
Y_i^A n_i=0,\quad Y_i^A=-\nabla^An_i.
\ee  For the vector $\widetilde{Y}_i^A$, we also find 
\be 
\widetilde{Y}_i^A n_i=0.
\ee 
\item Inner product. The inner product of $Y_{ij}^A$ and $Y_k^A$ is 
\be 
Y_{ij}\cdot Y_k=\delta_{ik}n_j-\delta_{jk}n_{i}.
\ee 
\item There are also some useful identities {involving} the Levi-Civita tensor 
\bea 
&& Y_i^AY_j^B-Y_i^BY_j^A=\epsilon^{AB}\epsilon_{ijk}n_k,\\
&& \gamma_{AB}{\widetilde{Y}_i^A Y_j^B}=\epsilon_{ijk}n_k,\\
&& \epsilon^{AB}\epsilon_{ijk}Y_k^C=-Y_{ij}^A\gamma^{BC}+Y_{ij}^B\gamma^{AC},\\
&&\frac{1}{2}\epsilon^{AB}\epsilon_{ijk}Y_{jk}^C=-\gamma^{AC}Y_i^B+\gamma^{BC}Y_i^A.
\eea 


\end{enumerate}
\subsection{Smooth vector fields}\label{smvsec}
We collect several properties which are used in the context. \begin{itemize}
\item 
For general smooth vectors $Y^A,Z^A$ on $S^2$, we have 
\bea 
Y^A\nabla_A\nabla_B Z^B-Z^B\nabla_B\nabla_AY^A=\nabla_A[Y,Z]^A.
\eea  This identity can be proved by using the commutator
\bea 
\ [\nabla_A,\nabla_B]V^C=R^C_{\ DAB}V^D\label{comuRie}
\eea where $V^C$ is any vector field on $S^2$. The Riemann tensor {in} \eqref{comuRie} is 
\be 
R_{ABCD}=\gamma_{AC}\gamma_{BD}-\gamma_{AD}\gamma_{BC}.
\ee The corresponding Ricci tensor and Ricci scalar are 
\be 
R_{AB}=\gamma_{AB},\quad R=2.
\ee 
\item For any two smooth vectors $Y^A,Z^A$  and smooth function $g$ on $S^2$, there is an identity 
\bea 
[Y,Z]^A\nabla_A g=Y^A\nabla_AZ^B\nabla_B g-Z^A\nabla_AY^B\nabla_B g.
\eea This identity is useful for checking the Jacobi identity 
\bea 
\ [\mathcal{M}_Y,[\mathcal{M}_Z,\mathcal{O}_g]]+[\mathcal{M}_{Z},[\mathcal{O}_g,\mathcal{M}_Y]]+[\mathcal{O}_g,[\mathcal{M}_Y,\mathcal{M}_Z]]=0.
\eea 
\item For any smooth vector $Y^A$ on $S^2$, we can construct a symmetric traceless tensor 
\be 
\Theta_{AB}(Y)=\nabla_AY_B+\nabla_BY_A-\gamma_{AB}\nabla_CY^C,\quad \Theta_{AB}(Y)=\Theta_{BA}(Y),\quad \Theta_A^{\ A}(Y)=0.
\ee There is a Fierz identity related to this tensor 
\be 
\epsilon_{A}^{\ B}\Theta_{BC}(Y)-\epsilon_{C}^{\ B}\Theta_{BA}(Y)=0.
\ee 
The function $\Theta_{AB}(Y)$ is related to the Lie derivative by\footnote{ Here the Lie derivative is defined on the unit sphere
\be 
\mathcal{L}_Y\gamma_{AB}=\nabla_AY_B+\nabla_BY_A,\nn
\ee while the one used in the context is defined on the Minkowski spacetime.}
\bea  
\Theta_{AB}(Y)&=&\mathcal{L}_Y\gamma_{AB}-\gamma_{AB}\nabla_CY^C,
\eea  We also find 
\bea  
\nabla^A\nabla^B\Theta_{AB}=\Box\nabla\cdot Y+2\nabla\cdot Y.
\eea
\item For any three smooth vectors $X^A,Y^A,Z^A$ on $S^2$, we can find the following identity 
\be 
X^A\nabla_Ao(Y,Z)+Y^A\nabla_Ao(Z,X)+Z^A\nabla_Ao(X,Y)+o(X,[Y,Z])+o(Y,[Z,X])+o(Z,[X,Y])=0.
\ee The function $o(Y,Z)$ is defined in \eqref{functiono}. This identity is useful to check the Jacobi identity 
\be 
\ [\mathcal{M}_X,[\mathcal{M}_Y,\mathcal{M}_Z]]+[\mathcal{M}_Y,[\mathcal{M}_Z,\mathcal{M}_X]]+[\mathcal{M}_Z,[\mathcal{M}_X,\mathcal{M}_Y]]=0.
\ee 
\end{itemize}

\section{Properties of the tensor \texorpdfstring{$P_{ABCD}$}{}}\label{rank4subsec}
The properties of the rank 4 tensor $P_{ABCD}$ are collected in the following. 
\begin{itemize}
    \item Symmetries
    \bea 
P_{ABCD}=P_{BADC}=P_{BDAC}=P_{DBCA}=P_{CDAB}=P_{DCBA}=P_{ACBD}=P_{CADB}.
\eea \item Traces
\bea 
P^A_{\ ABC}=P^A_{\ BAC}=P_{BC\hspace{5pt}A}^{\hspace{0.4cm} A}=2\gamma_{BC},\quad P^A_{\ BCA}=P_{B\hspace{3pt}AC}^{\ A}=0.
\eea 
\item Fierz identity
\be  
\epsilon^E_{\ B}P_{AECD}+\epsilon^E_{\ D}P_{ABCE}=0.
\ee  This identity follows  from the Fierz identity 
\be  
\epsilon_{AB}\gamma_{CD}+\epsilon_{BC}\gamma_{AD}+\epsilon_{CA}\gamma_{BD}=0.
\ee  
\item Product
\be 
\frac{1}{2}P_{ABCD}P_{E\hspace{3pt}F}^{\ B\hspace{3pt}D}=P_{ACEF}.
\ee 
\item The tensor $P_{ABCD}$ can also be written as 
\be 
P_{ABCD}=\gamma_{AC}\gamma_{BD}+\epsilon_{AC}\epsilon_{BD}.
\ee As a consequence, one can find
\bea 
P_{ABCD}+P_{ADCB}&=&2\gamma_{AC}\gamma_{BD},\\
P_{ABCD}-P_{ADCB}&=&2(\gamma_{AB}\gamma_{CD}-\gamma_{AD}\gamma_{BC})=2\epsilon_{AC}\epsilon_{BD}.
\eea 
\end{itemize}

\section{Canonical quantization}\label{canoquan} 
In perturbative quantum field theory, by imposing the Lorenz gauge 
\be 
\partial_\mu a^\mu=0,\label{lorgauge}
\ee the electromagnetic field $a_\mu$ may be quantized   using annihilation and creation operators $b_{\alpha,\bm k}, b^\dagger_{\alpha,\bm k}$ 
\bea 
a_\mu(t,\bm x)&=&\sum_{\alpha=\pm}\int \frac{d^3\bm k}{(2\pi)^3}\frac{1}{\sqrt{2\omega_{\bm k}}}[\epsilon^{*\alpha}_\mu(\bm k)b_{\alpha,\bm k}e^{-i\omega t+i\bm k\cdot\bm x}+\epsilon^{\alpha}_\mu(\bm k)b^\dagger_{\alpha,\bm k}e^{i\omega t-i\bm k\cdot\bm x}],\label{modeexpansion}
\eea  where the vector $\bm k$ is the momentum and $\omega$ is the energy of the corresponding mode. For a massless particle, we have
\be 
\omega=|\bm k|.
\ee The indices $\mu$ denote the spacetime coordinates and $\alpha$ is the polarization index.  The polarization vector  $\epsilon_\mu^\alpha(\bm k)$ has two physical degrees of freedom
\be 
\alpha=\pm.
\ee 
They also satisfy the completeness relation
\be 
\sum_{\alpha,\beta}\epsilon_{\mu}^{*\alpha}(\bm k)\delta_{\alpha,\beta}\epsilon_\nu^{\beta}(\bm k)=\bar{\eta}_{\mu\nu},
\ee where
\bea 
\bar{\eta}_{\mu\nu}=\eta_{\mu\nu}-\frac{1}{2}[n_\mu(\bm k)\bar{n}_\nu(\bm k)+n_\nu(\bm k)\bar{n}_{\mu}(\bm k)],
\eea 
with the normal vectors in momentum space
\begin{align}
    n_\mu(\bm k)=(-1,\frac{k_i}{|\bm k|}),\qquad \bar{n}_\mu(\bm k)=(1,\frac{k_i}{|\bm k|}).
\end{align}
The annihilation and creation operators satisfy the standard commutation relations
\bea 
\ [b_{\alpha,\bm k},b_{\beta,\bm k'}]&=&0,\\
\ [b_{\alpha,\bm k},b^\dagger_{\beta,\bm k'}]&=&(2\pi)^3\delta_{\alpha,\beta}\delta^{(3)}(\bm k-\bm k'),\\
\ [b_{\alpha,\bm k}^\dagger,b_{\beta,\bm k'}^\dagger]&=&0.
\eea  By expanding the plane wave into spherical waves,  the 
propagating modes $A_A$ are 
\bea 
A_A(u,\Omega)&=&\int_0^\infty \frac{d\omega}{\sqrt{4\pi\omega}}\sum_{\ell m}[c_{i;\omega,\ell,m}Y^i_A(\Omega)Y_{\ell,m}(\Omega)e^{-i\omega u}+c^\dagger_{i;\omega,\ell,m}Y^i_A(\Omega)Y_{\ell,m}^*(\Omega)e^{i\omega u}]\label{scriAA}
\eea with 
\bea 
c_{i;\omega,\ell,m}&=&\frac{i\omega}{(2\pi)^{3/2}}\int d\Omega_k\sum_{\alpha=\pm} \epsilon_i^{*\alpha}(\bm k)b_{\alpha,\bm k} Y_{\ell,m}^*(\Omega_k), \label{cannihilation}\\
c^\dagger_{i;\omega,\ell,m}&=&\frac{\omega}{(2\pi)^{3/2}i}\int d\Omega_k\sum_{\alpha=\pm} \epsilon_i^{\alpha}(\bm k)b^\dagger_{\alpha,\bm k} Y_{\ell,m}(\Omega_k).\label{ccreation}
\eea 
Their commutators are 
\bea 
\ [c_{i;\omega,\ell,m},c_{i';\omega',\ell',m'}]&=&0,\\
\ [c_{i;\omega,\ell,m},c^\dagger_{i';\omega',\ell',m'}]&=&\delta(\omega-\omega')\int d\Omega (\delta_{i,i'}-n_i n_{i'})Y^*_{\ell,m}(\Omega)Y_{\ell',m'}(\Omega),\label{cc}\\
\ [c^\dagger_{i;\omega,\ell,m},c^\dagger_{i';\omega',\ell',m'}]&=&0.
\eea
Then we find the  commutators 
at $\mathcal{I}^+$
\bea 
\ [A_A(u,\Omega),A_{B}(u',\Omega')]&=&\frac{i}{2}\gamma_{AB}\alpha(u-u')\delta(\Omega-\Omega'),\\
\ [A_{A}(u,\Omega),\dot A_{B}(u',\Omega')]&=&\frac{i}{2}\gamma_{AB}\delta(u-u')\delta(\Omega-\Omega'),\\
\ [\dot A_{A}(u,\Omega),\dot A_{B}(u',\Omega')]&=&\frac{i}{2}\gamma_{AB}\delta'(u-u')\delta(\Omega-\Omega').
\eea 
Notice that the second term on the right-hand side of \eqref{cc} will not contribute to the commutators due to $n_iY^A_i=0$. In fact, the first term is just the usual $\delta(\omega-\omega')\delta_{\ell,\ell'}\delta_{m,m'} $. We therefore get the desired commutators.

\section{Commutators}\label{comsec}
We will show the derivation of the commutators \eqref{LL2}-\eqref{OO}. Take the commutator $[\mathcal{M}_Y,\mathcal{M}_Z]$ as an example. We rewrite the superrotation generator as 
\bea 
\mathcal{M}_Y=\int du d\Omega \dot A^B\Delta_B(Y;A;u,\Omega).
\eea Here $\Delta_B(Y;A;u,\Omega)$ can be regarded as a smooth vector field on $S^2$. In the following, we use the abbreviation 
\be 
\Delta_B(Y;V)=\Delta_B(Y;V;u,\Omega)
\ee for any vector field $V$ on $S^2$. Then 
\bea 
\ [\mathcal{M}_Y,\mathcal{M}_Z]&=&\text{Central terms}+i\int du d\Omega \dot A^B[\Delta_B(Y;\Delta(Z;A))-\Delta_B(Z;\Delta(Y;A))]\nn\\&&+\text{Non-local terms}.
\eea 
 The central terms and the non-local terms can be calculated straightforwardly. The local terms are found by using the identity 
 \bea 
 \Delta_B(Y;\Delta(Z;A))-\Delta_B(Z;\Delta(Y;A))=\Delta_B([Y,Z];A)+o(Y,Z)\epsilon_{DB}A^D.
 \eea To prove it, we rewrite 
 \bea 
 \Delta_B(Y;A)=Y^C\nabla_CA_B+K_{BC}(Y)A^C,
 \eea where 
 \bea 
 K_{BC}(Y)=\nabla_BY_C-\frac{1}{2}\Theta_{BC}(Y).\label{KBC}
 \eea 
 Then 
 \bea 
&& \Delta_B(Y;\Delta(Z;A))-\Delta_B(Z;\Delta(Y;A))-\Delta_B([Y,Z];A)\nn\\&=&\ [Y^C\nabla_C\Delta_B(Z;A)+K_{BC}(Y)\Delta^C(Z;A)]-(Y\leftrightarrow Z)-[Y,Z]^C\nabla_CA_B-K_{BC}([Y,Z])A^C\nn\\&=&Y^C\nabla_C(Z^D\nabla_DA_B+K_{BD}(Z)A^D)+K_{B}^{\ C}(Y)(Z^D\nabla_D A_C+K_{CD}(Z)A^D)\nn\\&&-(Y\leftrightarrow Z)-[Y,Z]^C\nabla_CA_B-K_{BC}([Y,Z])A^C\nn\\&=&A^DL_{BD},
 \eea where 
 \bea 
 L_{BD}&=&[Y_BZ_D+Y^C\nabla_CK_{BD}(Z)+K_B^{\ C}(Y)K_{CD}(Z)]-(Y\leftrightarrow Z)-K_{BD}([Y,Z]).
 \eea Using the definition \eqref{KBC}, we find 
 \bea
 L_{BD}&=&[-\frac{1}{2}Y^C\nabla_C \Theta_{BD}(Z)-\frac{1}{2}\nabla^CZ_D\Theta_{BC}(Y)-\frac{1}{2}\nabla_BY_C \Theta^C_{\ D}(Z)+\frac{1}{4}\Theta^C_{\ D}(Z)\Theta_{BC}(Y)]\nn\\&&-(Y\leftrightarrow Z)+\frac{1}{2}\Theta_{BD}([Y,Z])\nn\\&=&[-\frac{1}{2}\mathcal{L}_Y \Theta_{BD}(Z)+\frac{1}{2}(\mathcal{L}_Y \gamma_{CD})\Theta_B^{\ C}(Z)+\frac{1}{4}\Theta^C_{\ D}(Z)\Theta_{BC}(Y)]-(Y\leftrightarrow Z)\nn\\&&+\frac{1}{2}\mathcal{L}_{[Y,Z]}\gamma_{BD}-\frac{1}{2}\gamma_{BD}\nabla_C[Y,Z]^C\nn\\&=&[-\frac{1}{2}\mathcal{L}_Y\mathcal{L}_Z\gamma_{BD}+\frac{1}{2}\mathcal{L}_Y(\gamma_{BD}\nabla_CZ^C)+\frac{1}{4}\Theta_{CD}(Y)\Theta_{B}^{\ C}(Z)+\frac{1}{2}\Theta_{BD}(Z)\nabla_CY^C]-(Y\leftrightarrow Z)\nn\\&&+\frac{1}{2}\mathcal{L}_{[Y,Z]}\gamma_{BD}-\frac{1}{2}\gamma_{BD}\nabla_C[Y,Z]^C\nn\\&=&\frac{1}{4}\Theta_{CD}(Y)\Theta_{B}^{\ C}(Z)-(Y\leftrightarrow Z)\nn\\&=&o(Y,Z)\epsilon_{DB}.
 \eea  
Therefore, the local terms in $[\mathcal{M}_{Y},\mathcal{M}_Z]$ are 
\bea 
\text{Local terms of}\ [\mathcal{M}_{Y},\mathcal{M}_Z]=i\mathcal{M}_{[Y,Z]}+i\mathcal{O}_{o(Y,Z)}.
\eea 
 
\section{Green's functions}\label{greensec}
The antipodal matching condition can also be checked using Green's functions of Maxwell equation 
\be 
\partial_\mu f^{\mu\nu}=-j^\nu.
\ee The vector potential can be solved in Lorenz gauge using retarded Green's function 
\bea 
a_\mu(t,\bm x)&=&a^{\text{in}}_\mu(t,\bm x)+a^{\text{ret}}_\mu(t,\bm x),
\eea where the retarded solution is 
\bea 
a^{\text{ret}}_\mu (t,\bm x)=\int d\bm x' \frac{j_\mu(t-|\bm x-\bm x'|)}{4\pi|\bm x-\bm x'|}.
\eea The ingoing wave $a^{\text{in}}(t,\bm x)$ is determined by imposing the initial conditions at $\mathcal{I}^-$.  The vector potential can also be represented in terms of advanced Green's function 
\bea 
a_\mu(t,\bm x)&=&a^{\text{out}}_\mu(t,\bm x)+a^{\text{adv}}_\mu(t,\bm x).
\eea The advanced solution is 
\bea 
a^{\text{adv}}_\mu(t,\bm x)=\int d\bm x' \frac{j_\mu(t+|\bm x-\bm x'|)}{4\pi|\bm x-\bm x'|}
\eea and the outgoing wave is denoted as $a^{\text{out}}_\mu(t,\bm x)$. The radiation field is the difference between the outgoing wave and the ingoing wave \cite{Dirac:1938nz}
\bea 
a_\mu^{\text{rad}}(t,\bm x)&=&a^{\text{out}}_\mu(t,\bm x)-a^{\text{in}}_\mu(t,\bm x)=a^{\text{ret}}_\mu(t,\bm x)-a^{\text{adv}}_\mu(t,\bm x).
\eea 
Using the Fourier transformation 
\bea 
j_\mu(t,\bm x)=\int \frac{d\omega d^3\bm k}{(2\pi)^4}e^{-i\omega t+i\bm k\cdot \bm x}j_\mu(\omega,\bm k),
\eea the leading term of the vector field near $\mathcal{I}^+$ is 
\bea 
A_\mu(u,\Omega)&=&\frac{1}{8\pi^2}\int d\omega e^{-i\omega u}j_\mu(\omega,\bm k),\quad \bm k=(\omega,\Omega).
\eea In Fourier space, we find 
\bea
A_\mu(\omega,\Omega)=\frac{j_\mu(\omega,\bm k)}{4\pi}.
\eea 
In spherical coordinates, we find 
\bea 
A_A(u,\Omega)&=&-Y^i_AA_i(u,\Omega)=-\frac{1}{8\pi^2}\int d\omega e^{-i\omega u}j_i(\omega,\bm k)Y^i_A(\Omega),\quad \bm k=(\omega,\Omega).
\eea 
Similarly, we also find 
\bea 
A^-_\mu(v,\Omega)&=&-\frac{1}{8\pi^2}\int d\omega e^{-i\omega v}j_\mu(\omega,-\bm k),\quad \bm k=(\omega,\Omega)
\eea and 
\be 
A^-_\mu(\omega,\Omega)=-\frac{j_\mu(\omega,-\bm k)}{4\pi}.
\ee 
In advanced coordinates,  
\bea 
A^-_A(v,\Omega)&=&-Y^i_A A_i^-(v,\Omega)=\frac{1}{8\pi^2}\int d\omega e^{-i\omega v}j_i(\omega,-\bm k)Y^i_A(\Omega),\quad \bm k=(\omega,\Omega).
\eea In spherical coordinates, the $-\bm k$ is
\be 
-\bm k=(\omega,\Omega^P),
\ee where $\Omega^P$ is the antipodal point of $\Omega$
\be 
\Omega^P=(\pi-\theta,\pi+\phi).
\ee There is an antipodal matching condition in frequency space
\be 
A_\mu(\omega,\Omega)=-A_\mu^-(\omega,\Omega^P).
\ee 
The radiation electric and magnetic fields are 
\bea 
E_i(u,\Omega)&=&Y_i^A\dot A_A(u,\Omega)=-(\delta_{ij}-n_in_j)\dot A_{ j}(u,\Omega),\\
B_i(u,\Omega)&=&-\widetilde{Y}_i^A \dot A_A(u,\Omega)=\epsilon_{ijk}n_k \dot A_j(u,\Omega),\\
E_i^-(v,\Omega)&=&Y_i^A\dot A_A^-(v,\Omega)=-(\delta_{ij}-n_in_j)\dot A_j^-(v,\Omega),\\
B_i^-(v,\Omega)&=&\widetilde{Y}_i^A\dot A_A^-(v,\Omega)=-\epsilon_{ijk}n_k \dot A_j^-(v,\Omega).
\eea 
In frequency  space, they are
\bea 
E_i(\omega,\Omega)&=&i\omega (\delta_{ij}-n_in_j)A_j(\omega,\Omega),\\
B_i(\omega,\Omega)&=&-i\epsilon_{ijk}n_k A_j(\omega,\Omega),\\
E_i^-(\omega,\Omega)&=&i\omega (\delta_{ij}-n_in_j)A^-_j(\omega,\Omega),\\
B_i^-(\omega,\Omega)&=&i\omega \epsilon_{ijk}n_k A_j^-(\omega,\Omega).
\eea 
Since 
\be 
n_i(\Omega^P)=-n_i(\Omega),
\ee the antipodal matching conditions for the electric and magnetic fields are
\bea 
E_i(\omega,\Omega)=-E_i^-(\omega,\Omega^P),\quad B_i(\omega,\Omega)=-B_i^-(\omega,\Omega^P).
\eea By expanding the Green{'s} function to order $\mathcal{O}(r^{-2})$, we can also find the antipodal matching condition at the subleading order
\bea 
E^{(2)}_i(\omega,\Omega)=E_i^{-(2)}(\omega,\Omega^P),\quad B_i^{(2)}(\omega,\Omega)=B_i^{-(2)}(\omega,\Omega^P).
\eea 
\bibliography{refs}

\providecommand{\href}[2]{#2}\begingroup\raggedright\begin{thebibliography}{10}

\bibitem{Bondi:1962px}
H.~Bondi, M.~G.~J. van~der Burg, and A.~W.~K. Metzner, ``{Gravitational waves
  in general relativity. 7. Waves from axisymmetric isolated systems},'' {\em
  Proc. Roy. Soc. Lond. A} {\bf 269} (1962) 21--52.

\bibitem{Sachs:1962wk}
R.~K. Sachs, ``{Gravitational waves in general relativity. 8. Waves in
  asymptotically flat space-times},'' {\em Proc. Roy. Soc. Lond. A} {\bf 270}
  (1962) 103--126.

\bibitem{Sachs:1962zza}
R.~Sachs, ``{Asymptotic symmetries in gravitational theory},'' {\em Phys. Rev.}
  {\bf 128} (1962) 2851--2864.

\bibitem{Barnich:2010eb}
G.~Barnich and C.~Troessaert, ``{Aspects of the BMS/CFT correspondence},'' {\em
  JHEP} {\bf 05} (2010) 062, \href{http://www.arXiv.org/abs/1001.1541}{{\tt
  1001.1541}}.

\bibitem{Barnich:2009se}
G.~Barnich and C.~Troessaert, ``{Symmetries of asymptotically flat 4
  dimensional spacetimes at null infinity revisited},'' {\em Phys. Rev. Lett.}
  {\bf 105} (2010) 111103, \href{http://www.arXiv.org/abs/0909.2617}{{\tt
  0909.2617}}.

\bibitem{Barnich:2010ojg}
G.~Barnich and C.~Troessaert, ``{Supertranslations call for superrotations},''
  {\em PoS} {\bf CNCFG2010} (2010) 010,
  \href{http://www.arXiv.org/abs/1102.4632}{{\tt 1102.4632}}.

\bibitem{Barnich:2011mi}
G.~Barnich and C.~Troessaert, ``{BMS charge algebra},'' {\em JHEP} {\bf 12}
  (2011) 105, \href{http://www.arXiv.org/abs/1106.0213}{{\tt 1106.0213}}.

\bibitem{Campiglia:2014yka}
M.~Campiglia and A.~Laddha, ``{Asymptotic symmetries and subleading soft
  graviton theorem},'' {\em Phys. Rev. D} {\bf 90} (2014), no.~12, 124028,
  \href{http://www.arXiv.org/abs/1408.2228}{{\tt 1408.2228}}.

\bibitem{Campiglia:2015yka}
M.~Campiglia and A.~Laddha, ``{New symmetries for the Gravitational
  S-matrix},'' {\em JHEP} {\bf 04} (2015) 076,
  \href{http://www.arXiv.org/abs/1502.02318}{{\tt 1502.02318}}.

\bibitem{Strominger:2013jfa}
A.~Strominger, ``{On BMS Invariance of Gravitational Scattering},'' {\em JHEP}
  {\bf 07} (2014) 152, \href{http://www.arXiv.org/abs/1312.2229}{{\tt
  1312.2229}}.

\bibitem{Strominger:2017zoo}
A.~Strominger, ``{Lectures on the Infrared Structure of Gravity and Gauge
  Theory},'' \href{http://www.arXiv.org/abs/1703.05448}{{\tt 1703.05448}}.

\bibitem{Kapec:2016jld}
D.~Kapec, P.~Mitra, A.-M. Raclariu, and A.~Strominger, ``{2D Stress Tensor for
  4D Gravity},'' {\em Phys. Rev. Lett.} {\bf 119} (2017), no.~12, 121601,
  \href{http://www.arXiv.org/abs/1609.00282}{{\tt 1609.00282}}.

\bibitem{Pasterski:2016qvg}
S.~Pasterski, S.-H. Shao, and A.~Strominger, ``{Flat Space Amplitudes and
  Conformal Symmetry of the Celestial Sphere},'' {\em Phys. Rev. D} {\bf 96}
  (2017), no.~6, 065026, \href{http://www.arXiv.org/abs/1701.00049}{{\tt
  1701.00049}}.

\bibitem{Pasterski:2017kqt}
S.~Pasterski and S.-H. Shao, ``{Conformal basis for flat space amplitudes},''
  {\em Phys. Rev. D} {\bf 96} (2017), no.~6, 065022,
  \href{http://www.arXiv.org/abs/1705.01027}{{\tt 1705.01027}}.

\bibitem{Raclariu:2021zjz}
A.-M. Raclariu, ``{Lectures on Celestial Holography},''
  \href{http://www.arXiv.org/abs/2107.02075}{{\tt 2107.02075}}.

\bibitem{Pasterski:2021rjz}
S.~Pasterski, ``{Lectures on celestial amplitudes},'' {\em Eur. Phys. J. C}
  {\bf 81} (2021), no.~12, 1062,
  \href{http://www.arXiv.org/abs/2108.04801}{{\tt 2108.04801}}.

\bibitem{Donnay:2022aba}
L.~Donnay, A.~Fiorucci, Y.~Herfray, and R.~Ruzziconi, ``{Carrollian Perspective
  on Celestial Holography},'' {\em Phys. Rev. Lett.} {\bf 129} (2022), no.~7,
  071602, \href{http://www.arXiv.org/abs/2202.04702}{{\tt 2202.04702}}.

\bibitem{Donnay:2022wvx}
L.~Donnay, A.~Fiorucci, Y.~Herfray, and R.~Ruzziconi, ``{Bridging Carrollian
  and Celestial Holography},'' \href{http://www.arXiv.org/abs/2212.12553}{{\tt
  2212.12553}}.

\bibitem{Duval_2014a}
C.~Duval, G.~W. Gibbons, and P.~A. Horvathy, ``Conformal carroll groups and
  {BMS} symmetry,'' {\em Classical and Quantum Gravity} {\bf 31} (apr, 2014)
  092001.

\bibitem{Duval_2014b}
C.~Duval, G.~W. Gibbons, and P.~A. Horvathy, ``Conformal carroll groups,'' {\em
  Journal of Physics A: Mathematical and Theoretical} {\bf 47} (aug, 2014)
  335204.

\bibitem{Duval:2014uoa}
C.~Duval, G.~W. Gibbons, P.~A. Horvathy, and P.~M. Zhang, ``{Carroll versus
  Newton and Galilei: two dual non-Einsteinian concepts of time},'' {\em Class.
  Quant. Grav.} {\bf 31} (2014) 085016,
  \href{http://www.arXiv.org/abs/1402.0657}{{\tt 1402.0657}}.

\bibitem{Une}
J.~M. L\'evy-Leblond, ``{Une nouvelle limite non-relativiste du groupe de
  Poincar\'e},'' {\em Ann. Inst. H Poincar\'e} {\bf 3} (1965), no.~1, 1--12.

\bibitem{Gupta1966OnAA}
N.~Gupta, ``On an analogue of the galilei group,'' {\em Nuovo Cimento Della
  Societa Italiana Di Fisica A-nuclei Particles and Fields} {\bf 44} (1966)
  512--517.

\bibitem{Henneaux:1979vn}
M.~Henneaux, ``{Geometry of Zero Signature Space-times},'' {\em Bull. Soc.
  Math. Belg.} {\bf 31} (1979) 47--63.

\bibitem{Chen:2021xkw}
B.~Chen, R.~Liu, and Y.-f. Zheng, ``{On Higher-dimensional Carrollian and
  Galilean Conformal Field Theories},''
  \href{http://www.arXiv.org/abs/2112.10514}{{\tt 2112.10514}}.

\bibitem{Chen:2023pqf}
B.~Chen, R.~Liu, H.~Sun, and Y.-f. Zheng, ``{Constructing Carrollian Field
  Theories from Null Reduction},''
  \href{http://www.arXiv.org/abs/2301.06011}{{\tt 2301.06011}}.

\bibitem{Bagchi:2010zz}
A.~Bagchi, ``{Correspondence between Asymptotically Flat Spacetimes and
  Nonrelativistic Conformal Field Theories},'' {\em Phys. Rev. Lett.} {\bf 105}
  (2010) 171601, \href{http://www.arXiv.org/abs/1006.3354}{{\tt 1006.3354}}.

\bibitem{Bagchi:2016bcd}
A.~Bagchi, R.~Basu, A.~Kakkar, and A.~Mehra, ``{Flat Holography: Aspects of the
  dual field theory},'' {\em JHEP} {\bf 12} (2016) 147,
  \href{http://www.arXiv.org/abs/1609.06203}{{\tt 1609.06203}}.

\bibitem{Bagchi:2019xfx}
A.~Bagchi, A.~Mehra, and P.~Nandi, ``{Field Theories with Conformal Carrollian
  Symmetry},'' {\em JHEP} {\bf 05} (2019) 108,
  \href{http://www.arXiv.org/abs/1901.10147}{{\tt 1901.10147}}.

\bibitem{Bagchi:2019clu}
A.~Bagchi, R.~Basu, A.~Mehra, and P.~Nandi, ``{Field Theories on Null
  Manifolds},'' {\em JHEP} {\bf 02} (2020) 141,
  \href{http://www.arXiv.org/abs/1912.09388}{{\tt 1912.09388}}.

\bibitem{Banerjee:2020qjj}
K.~Banerjee, R.~Basu, A.~Mehra, A.~Mohan, and A.~Sharma, ``{Interacting
  Conformal Carrollian Theories: Cues from Electrodynamics},'' {\em Phys. Rev.
  D} {\bf 103} (2021), no.~10, 105001,
  \href{http://www.arXiv.org/abs/2008.02829}{{\tt 2008.02829}}.

\bibitem{Hao:2021urq}
P.-x. Hao, W.~Song, X.~Xie, and Y.~Zhong, ``{BMS-invariant free scalar
  model},'' {\em Phys. Rev. D} {\bf 105} (2022), no.~12, 125005,
  \href{http://www.arXiv.org/abs/2111.04701}{{\tt 2111.04701}}.

\bibitem{Henneaux:2021yzg}
M.~Henneaux and P.~Salgado-Rebolledo, ``{Carroll contractions of
  Lorentz-invariant theories},'' {\em JHEP} {\bf 11} (2021) 180,
  \href{http://www.arXiv.org/abs/2109.06708}{{\tt 2109.06708}}.

\bibitem{Bagchi:2022owq}
A.~Bagchi, D.~Grumiller, and P.~Nandi, ``{Carrollian superconformal theories
  and super BMS},'' {\em JHEP} {\bf 05} (2022) 044,
  \href{http://www.arXiv.org/abs/2202.01172}{{\tt 2202.01172}}.

\bibitem{Bagchi:2022xug}
A.~Bagchi, R.~Chatterjee, R.~Kaushik, S.~Pal, M.~Riegler, and D.~Sarkar.~a,
  ``{BMS Field Theories with $\mathfrak{u}(1)$ Symmetry},''
  \href{http://www.arXiv.org/abs/2209.06832}{{\tt 2209.06832}}.

\bibitem{Bekaert:2022ipg}
X.~Bekaert and B.~Oblak, ``{Massless Scalars and Higher-Spin BMS in Any
  Dimension},'' \href{http://www.arXiv.org/abs/2209.02253}{{\tt 2209.02253}}.

\bibitem{Rivera-Betancour:2022lkc}
D.~Rivera-Betancour and M.~Vilatte, ``{Revisiting the Carrollian scalar
  field},'' {\em Phys. Rev. D} {\bf 106} (2022), no.~8, 085004,
  \href{http://www.arXiv.org/abs/2207.01647}{{\tt 2207.01647}}.

\bibitem{Schwarz:2022dqf}
J.~H. Schwarz, ``{Diffeomorphism Symmetry in Two Dimensions and Celestial
  Holography},'' \href{http://www.arXiv.org/abs/2208.13304}{{\tt 2208.13304}}.

\bibitem{Dutta:2022vkg}
S.~Dutta, ``{Stress tensors of 3d Carroll CFTs},''
  \href{http://www.arXiv.org/abs/2212.11002}{{\tt 2212.11002}}.

\bibitem{Baiguera:2022lsw}
S.~Baiguera, G.~Oling, W.~Sybesma, and B.~T. S\o{}gaard, ``{Conformal Carroll
  Scalars with Boosts},'' \href{http://www.arXiv.org/abs/2207.03468}{{\tt
  2207.03468}}.

\bibitem{Bekaert:2022oeh}
X.~Bekaert, A.~Campoleoni, and S.~Pekar, ``{Carrollian conformal scalar as
  flat-space singleton},'' \href{http://www.arXiv.org/abs/2211.16498}{{\tt
  2211.16498}}.

\bibitem{Bagchi:2022eav}
A.~Bagchi, A.~Banerjee, S.~Dutta, K.~S. Kolekar, and P.~Sharma, ``{Carroll
  covariant scalar fields in two dimensions},''
  \href{http://www.arXiv.org/abs/2203.13197}{{\tt 2203.13197}}.

\bibitem{Saha:2022gjw}
A.~Saha, ``{Intrinsic approach to 1 + 1D Carrollian Conformal Field Theory},''
  {\em JHEP} {\bf 12} (2022) 133,
  \href{http://www.arXiv.org/abs/2207.11684}{{\tt 2207.11684}}.

\bibitem{Saha:2023hsl}
A.~Saha, ``{Carrollian Approach to $1+3$D Flat Holography},''
  \href{http://www.arXiv.org/abs/2304.02696}{{\tt 2304.02696}}.

\bibitem{Salzer:2023jqv}
J.~Salzer, ``{An Embedding Space Approach to Carrollian CFT Correlators for
  Flat Space Holography},'' \href{http://www.arXiv.org/abs/2304.08292}{{\tt
  2304.08292}}.

\bibitem{Liu:2022mne}
W.-B. Liu and J.~Long, ``{Symmetry group at future null infinity: scalar
  theory},'' {\em Phys. Rev. D} {\bf {107}} (2023) 126002,
  \href{http://www.arXiv.org/abs/2210.00516}{{\tt 2210.00516}}.

\bibitem{Ciambelli:2018xat}
L.~Ciambelli, C.~Marteau, A.~C. Petkou, P.~M. Petropoulos, and K.~Siampos,
  ``{Covariant Galilean versus Carrollian hydrodynamics from relativistic
  fluids},'' {\em Class. Quant. Grav.} {\bf 35} (2018), no.~16, 165001,
  \href{http://www.arXiv.org/abs/1802.05286}{{\tt 1802.05286}}.

\bibitem{Ciambelli:2019lap}
L.~Ciambelli, R.~G. Leigh, C.~Marteau, and P.~M. Petropoulos, ``{Carroll
  Structures, Null Geometry and Conformal Isometries},'' {\em Phys. Rev. D}
  {\bf 100} (2019), no.~4, 046010,
  \href{http://www.arXiv.org/abs/1905.02221}{{\tt 1905.02221}}.

\bibitem{Lee:1990nz}
J.~Lee and R.~M. Wald, ``{Local symmetries and constraints},'' {\em J. Math.
  Phys.} {\bf 31} (1990) 725--743.

\bibitem{Wald:1999wa}
R.~M. Wald and A.~Zoupas, ``{A General definition of 'conserved quantities' in
  general relativity and other theories of gravity},'' {\em Phys. Rev. D} {\bf
  61} (2000) 084027, \href{http://www.arXiv.org/abs/gr-qc/9911095}{{\tt
  gr-qc/9911095}}.

\bibitem{Ashtekar:1981bq}
A.~Ashtekar and M.~Streubel, ``{Symplectic Geometry of Radiative Modes and
  Conserved Quantities at Null Infinity},'' {\em Proc. Roy. Soc. Lond. A} {\bf
  376} (1981) 585--607.

\bibitem{Ashtekar:1981sf}
A.~Ashtekar, ``{Asymptotic Quantization of the Gravitational Field},'' {\em
  Phys. Rev. Lett.} {\bf 46} (1981) 573--576.

\bibitem{Ashtekar:1987tt}
A.~Ashtekar, {\em {Asymptotic Quantization: Based on 1984 Naples Lectures}}.
\newblock Bibliopolis, 1987.

\bibitem{1995iqft.book.....P}
M.~E. {Peskin} and D.~V. {Schroeder}, {\em {An Introduction to Quantum Field
  Theory}}.
\newblock {Addison-Wesley Pub}, 1995.

\bibitem{Oliver:1892}
H.~Oliver, ``Xi. on the forces, stresses, and fluxes of energy in the
  electromagnetic field,'' {\em Phil. Trans. R. Soc. A} {\bf 183} (1892)
  423--480.

\bibitem{Dirac:1931kp}
P.~A.~M. Dirac, ``{Quantised singularities in the electromagnetic field,},''
  {\em Proc. Roy. Soc. Lond. A} {\bf 133} (1931), no.~821, 60--72.

\bibitem{Dirac:1948um}
P.~A.~M. Dirac, ``{The Theory of magnetic poles},'' {\em Phys. Rev.} {\bf 74}
  (1948) 817--830.

\bibitem{1957AnPhy...2..525M}
C.~W. {Misner} and J.~A. {Wheeler}, ``{Classical physics as geometry},'' {\em
  Annals of Physics} {\bf 2} (Dec., 1957) 525--603.

\bibitem{Deser:1976iy}
S.~Deser and C.~Teitelboim, ``{Duality Transformations of Abelian and
  Nonabelian Gauge Fields},'' {\em Phys. Rev. D} {\bf 13} (1976) 1592--1597.

\bibitem{Dirac:1931mon}
P.A.M.Dirac, ``The quantum theory of the electron,'' {\em Proc. Roy. Soc. Lond.
  A} {\bf 133} (1931) 60.

\bibitem{Olive:1995sw}
D.~I. Olive, ``{Exact electromagnetic duality},'' {\em Nucl. Phys. B Proc.
  Suppl.} {\bf 45} (1996) 88--102,
  \href{http://www.arXiv.org/abs/hep-th/9508089}{{\tt hep-th/9508089}}.

\bibitem{Bliokh:2012zr}
K.~Y. Bliokh, A.~Y. Bekshaev, and F.~Nori, ``{Dual electromagnetism: Helicity,
  spin, momentum, and angular momentum},'' {\em New J. Phys.} {\bf 15} (2013)
  033026, \href{http://www.arXiv.org/abs/1208.4523}{{\tt 1208.4523}}.

\bibitem{2005math.ph...1052A}
S.~C. {Anco} and D.~{The}, ``{Symmetries, conservation laws, and cohomology of
  Maxwell's equations using potentials},'' {\em Acta Applicandae Mathematica}
  {\bf 89} (2005) 1--52, \href{http://www.arXiv.org/abs/math-ph/0501052}{{\tt
  math-ph/0501052}}.

\bibitem{Hosseinzadeh:2018dkh}
V.~Hosseinzadeh, A.~Seraj, and M.~M. Sheikh-Jabbari, ``{Soft Charges and
  Electric-Magnetic Duality},'' {\em JHEP} {\bf 08} (2018) 102,
  \href{http://www.arXiv.org/abs/1806.01901}{{\tt 1806.01901}}.

\bibitem{Henneaux:2020nxi}
M.~Henneaux and C.~Troessaert, ``{A note on electric-magnetic duality and soft
  charges},'' {\em JHEP} {\bf 06} (2020) 081,
  \href{http://www.arXiv.org/abs/2004.05668}{{\tt 2004.05668}}.

\bibitem{Maleknejad:2023nyh}
A.~Maleknejad, ``{Photon Chiral Memory Effect Stored on Celestial Sphere},''
  \href{http://www.arXiv.org/abs/2304.05381}{{\tt 2304.05381}}.

\bibitem{Oblak:2023axy}
B.~Oblak and A.~Seraj, ``{Orientation Memory of Magnetic Dipoles},''
  \href{http://www.arXiv.org/abs/2304.12348}{{\tt 2304.12348}}.

\bibitem{He:2014cra}
T.~He, P.~Mitra, A.~P. Porfyriadis, and A.~Strominger, ``{New Symmetries of
  Massless QED},'' {\em JHEP} {\bf 10} (2014) 112,
  \href{http://www.arXiv.org/abs/1407.3789}{{\tt 1407.3789}}.

\bibitem{Kapec:2014zla}
D.~Kapec, V.~Lysov, and A.~Strominger, ``{Asymptotic Symmetries of Massless QED
  in Even Dimensions},'' {\em Adv. Theor. Math. Phys.} {\bf 21} (2017)
  1747--1767, \href{http://www.arXiv.org/abs/1412.2763}{{\tt 1412.2763}}.

\bibitem{Kapec:2015ena}
D.~Kapec, M.~Pate, and A.~Strominger, ``{New Symmetries of QED},'' {\em Adv.
  Theor. Math. Phys.} {\bf 21} (2017) 1769--1785,
  \href{http://www.arXiv.org/abs/1506.02906}{{\tt 1506.02906}}.

\bibitem{Campiglia:2015qka}
M.~Campiglia and A.~Laddha, ``{Asymptotic symmetries of QED and
  Weinberg\textquoteright{}s soft photon theorem},'' {\em JHEP} {\bf 07} (2015)
  115, \href{http://www.arXiv.org/abs/1505.05346}{{\tt 1505.05346}}.

\bibitem{Kapec:2017tkm}
D.~Kapec, M.~Perry, A.-M. Raclariu, and A.~Strominger, ``{Infrared Divergences
  in QED, Revisited},'' {\em Phys. Rev. D} {\bf 96} (2017), no.~8, 085002,
  \href{http://www.arXiv.org/abs/1705.04311}{{\tt 1705.04311}}.

\bibitem{Nande:2017dba}
A.~Nande, M.~Pate, and A.~Strominger, ``{Soft Factorization in QED from 2D
  Kac-Moody Symmetry},'' {\em JHEP} {\bf 02} (2018) 079,
  \href{http://www.arXiv.org/abs/1705.00608}{{\tt 1705.00608}}.

\bibitem{Henneaux:2018gfi}
M.~Henneaux and C.~Troessaert, ``{Asymptotic symmetries of electromagnetism at
  spatial infinity},'' {\em JHEP} {\bf 05} (2018) 137,
  \href{http://www.arXiv.org/abs/1803.10194}{{\tt 1803.10194}}.

\bibitem{Freidel:2018fsk}
L.~Freidel and D.~Pranzetti, ``{Electromagnetic duality and central charge},''
  {\em Phys. Rev. D} {\bf 98} (2018), no.~11, 116008,
  \href{http://www.arXiv.org/abs/1806.03161}{{\tt 1806.03161}}.

\bibitem{Fuentealba:2023rvf}
O.~Fuentealba, M.~Henneaux, and C.~Troessaert, ``{A note on the asymptotic
  symmetries of electromagnetism},''
  \href{http://www.arXiv.org/abs/2301.05989}{{\tt 2301.05989}}.

\bibitem{Ferko:2021bym}
C.~Ferko, G.~Satishchandran, and S.~Sethi, ``{Gravitational memory and compact
  extra dimensions},'' {\em Phys. Rev. D} {\bf 105} (2022), no.~2, 024072,
  \href{http://www.arXiv.org/abs/2109.11599}{{\tt 2109.11599}}.

\bibitem{Dirac:1938nz}
P.~A.~M. Dirac, ``{Classical theory of radiating electrons},'' {\em Proc. Roy.
  Soc. Lond. A} {\bf 167} (1938) 148--169.

\end{thebibliography}\endgroup

\end{document}